\documentclass[pdflatex,sn-mathphys-num]{sn-jnl}

\usepackage{graphicx}%
\usepackage{multirow}%
\usepackage{amsmath,amssymb,amsfonts}%
\usepackage{amsthm}%
\usepackage{xcolor}%
\usepackage{textcomp}%
\usepackage{manyfoot}%
\usepackage{booktabs}%
\usepackage{algorithm}%
\usepackage{algorithmicx}%
\usepackage{algpseudocode}%
\usepackage{listings}%
\usepackage{subcaption}
\usepackage{float}
\usepackage{cite}

\theoremstyle{thmstyleone}%
\theoremstyle{thmstyletwo}%

\theoremstyle{thmstylethree}%

\begin{document}

\title[]{$\mathcal{PT}$-invariant nonlocal variable-coefficient
	Fokas-Lenells equation: nonstandard Hirota
	bilinearization, symmetry-preserving and breaking
	multi-soliton solutions}
	
\author*[1]{\fnm{Sagardeep} \sur{Talukdar}}\email{talukdarsagardeep@gmail.com}

\author[2]{\fnm{Ramakrishnan} \sur{R}}\email{ramakrishnan.cnld@gmail.com}

\author[1]{\fnm{Sudipta} \sur{Nandy}}\email{sudiptanandy@gmail.com}

\author[3]{\fnm{Lakshmanan} \sur{M}}\email{lakshman.cnld@gmail.com}

\affil[1]{\orgdiv{Department of Physics}, \orgname{Cotton University}, \orgaddress{\street{Panbazar}, \city{Guwahati}, \postcode{781001}, \state{Assam}, \country{India}}}

\affil[2]{\orgdiv{Department of Physics}, \orgname{Indian Institute of Technology Kharagpur}, \orgaddress{\street{}, \city{Kharagpur}, \postcode{721302}, \state{West Bengal}, \country{India}}}

\affil[3]{\orgdiv{Department of Nonlinear Dynamics}, \orgname{Bharathidasan University}, \orgaddress{\street{}, \city{Tiruchirappalli}, \postcode{620024}, \state{Tamil Nadu}, \country{India}}}



\abstract{In this paper, we explore a variable-coefficient Fokas-Lenells system with gain/loss. Applying a nonlocal symmetry reduction, the system leads to a $\mathcal{PT}$-invariant variable-coefficient Fokas-Lenells equation (vcFLE) with specific constraints on dispersion and nonlinearity coefficients. We prove the Lax integrability of the given system and construct one- and two-soliton solutions using a nonstandard Hirota bilinearization method. We show that, depending on the soliton parameters, the solutions can retain or violate the underlying symmetry. With appropriate choice of dispersion and nonlinearity, the system shows periodic-wave and soliton structures.}




\maketitle

\section{Introduction}
\label{sec1}
About a decade ago, a new type of integrable system was first proposed by Ablowitz and Musslimani \citep{ablowitz2013integrable}, namely the nonlinear Schr\"odinger equation (NLSE) of reverse space type, also known as the nonlocal NLSE with $\mathcal{PT}$-symmetry. 
Since then, studies on the properties of the $\mathcal{PT}$-symmetric NLSE have been extensively carried out in refs.\citep{ablowitz2016inverse,stalin2017nonstandard,chen2018solutions,gurses2018nonlocal,priya2019symmetry}. 
Unlike the local equations whose solution directly depends on  the local values of $x$ and $t$, the solution of the nonlocal integrable equations depends not only on $x$, $t$ but also on $-x$ and $-t$, that is, they are directly coupled. 
To be more specific, the solution of the local equation is evaluated in the space-time point $(x,t)$. However, in the nonlocal case, in addition to the point $(x,t)$, the solution is simultaneously evaluated at distant space-time points $(-x,t)$, $(x,-t)$ and/or $(-x,-t)$. Over the past ten years after the work in ref.\citep{ablowitz2013integrable}, there have been several studies on other nonlocal equations such as the nonlocal modified Korteweg-de Vries (nonlocal mKdV) equation \citep{gurses2019nonlocal}, etc., to investigate their properties. Moreover, Yang et al. proposed the localized wave solutions of the reverse space nonlocal Lakshmanan-Porsezian-Daniel equation \citep{yang2020darboux}. Apart from these, there are many works extensively carried out on various other nonlocal equations. Few of them are mentioned in refs. \citep{ablowitz2014integrable,yan2015integrable,ablowitz2017integrable,ablowitz2018reverse}. \\
\indent More recently, the study has also been extended to nonlocal FLE to find its exact soliton solution \citep{fokas2016integrable,zhang2022integrability}. The exact solutions of local FLE have been well studied through various analytical methods, namely the inverse scattering technique\citep{lashkin2021perturbation}, the Hirota bilinear method\citep{Matsuno1,Matsuno2}, etc.  Additionally, solutions of nonlocal FLE are also obtained by Darboux transformation \citep{zhang2019exact}, and the Riemann-Hilbert method\citep{li2021n}.  Zhang et al. in ref. \citep{zhang2022integrability} proposed a reverse space nonlocal FLE and showed that under suitable variable transformation the proposed system can transform into the  nonlocal reverse time and reverse space-time equations. Most of these works on the nonlocal FLE focus on constant dispersion as well as nonlinearity along the fiber. This opens up a question on how the nonlocal solitons arising out of various integrable systems behave when additional spatially/temporally varying dispersion or nonlinearity is introduced in the system. 

\indent So, in this paper we introduce the reverse space-time nonlocal variable-coefficient FLE (nonlocal vcFLE) as\citep{zhang2022integrability},
\begin{align}\label{cfle3}
	&q_{xt}(x,t) - M(t) \ q(x,t) -[2  N(t)\ q(x,t)\ q^*(-x,-t) +\Gamma(t)]q_x(x,t)   = 0, 
	\\ \label{cfle4}
	&q^*_{xt}(-x,-t) - M(t)\ q^*(-x,-t) +[2  N(t)\ q^*(-x,-t)\ q(x,t) -\Gamma(t)]q^*_x(-x,-t)   = 0,			
\end{align}

where $q(x,t)$ and $q^*(-x,-t)$ are the complex valued functions of an optical field at location $x$, $t$ and $-x$, $-t$, respectively. One novel approach in this manuscript is that we consider both $q(x,t)$ and $q^*(-x,-t)$ as separate field variables and not the complex conjugate of each other unlike the system considered in ref. \citep{zhang2022integrability}. This allows us to observe the spontaneous symmetry breaking in system (\ref{cfle3})-(\ref{cfle4}) and subsequently to obtain the symmetry preserving and breaking soliton solution for it. Moreover, in Eqs. (\ref{cfle3})-(\ref{cfle4}),  
$M(t)$ and $N(t)$ are the dispersion and nonlinearity coefficients respectively, and $\Gamma(t)=\frac{W[N(t), M(t)]}{2 M(t)N(t)}$ is the loss/gain term, where the wronskian $W[N(t), M(t)]=N(t)M'(t)-M(t)N'(t)$. The term $\Gamma(t)$ arises from the zero curvature equation and will be discussed in the next section.\\
\indent One can notice that the Eqs. (\ref{cfle3})-(\ref{cfle4}) are invariant under the joint action of $x\rightarrow-x$, $t\rightarrow-t$ and complex conjugation but under some condition imposed on the time varying coefficients. The conditions $M(-t)=M(t)$, $N(-t)=N(t)$ and $\Gamma(-t)=-\Gamma(t)$ ensure the $\mathcal{PT}$-symmetry of the system. This set of conditions restricts the dispersion and nonlinearity coefficients $M(t)$, $N(t)$ to be even function in time, and the gain/loss coefficient being their combination is an odd function in time.  The gain/loss parameter becomes non-zero ($\Gamma(t)$$\neq$0)
only if $M(t)$ and $N(t)$ are linearly independent. However, if they are linearly dependent, then $\Gamma(t)=0$. Under the special choice of the coefficients $M(t)=1$, $N(t)=1$, $\Gamma(t)$ becomes zero and the system (\ref{cfle3})-(\ref{cfle4}) reduces to the constant coefficient reverse space-time nonlocal FLE studied in ref.\citep{zhang2022integrability}.  \\
\indent The structure of the manuscript is as follows:. In Sec. \ref{section:2}, we introduce the Lax pair for the system (\ref{cfle3})-(\ref{cfle4}). Then in Sec. \ref{section:3}, we introduce the nonstandard Hirota bilinearization method and obtain a set of bilinear equations using auxiliary functions. Subsequently, solving the obtained linear partial differential equations (PDEs) from the bilinear equations, we find the one soliton, two soliton  and systematically discuss their properties and provide the scheme to obtain the $N$-soliton solution. Finally, in the last section we provide a brief conclusion. 

\section{Lax pair and integrability} 
\label{section:2}
\indent In this section, we construct the Lax pair $(L,M)$ for Eqs. (\ref{cfle3})-(\ref{cfle4}) and establish the integrability of the system. Equivalent linear equations for the variable-coefficient nonlocal FLE in terms of Lax pair is given by
\begin{align}
	\label{LaxFL}
	\partial_x { \Psi} = L { \Psi}, \\
	\partial_t { \Psi} = M { \Psi}.
\end{align}
Here, $\Psi$ is a two component vector field, expressed as,
\begin{align}
	\label{Psi}
	\Psi= ({\Psi_1} \ {\Psi_2})^T.
\end{align} 
Also, $L$ and $M$ are given by,
\begin{align}
	\label{Lax2}
	L=&\left(
	\begin{array}{cc}
		\frac{-i}{2}  \lambda ^2 & \lambda  q_x \sqrt{\frac{N(t)}{M(t)}} \\
		-\lambda  r_x \sqrt{\frac{N(t)}{M(t)}} & \frac{i}{2}\lambda ^2 \\
	\end{array}
	\right) ,     \\
	M=&\left(
	\begin{array}{cc}
		\frac{i M(t)}{2 \lambda ^2}-i r q N(t) & -\frac{i q M(t) \sqrt{\frac{N(t)}{M(t)}}}{\lambda } \\
		-\frac{i r M(t) \sqrt{\frac{N(t)}{M(t)}}}{\lambda } & i r q N(t)-\frac{i M(t)}{2 \lambda ^2} \\
	\end{array}
	\right),
\end{align} 
where $\lambda$ is the spectral parameter. The compatibility equation of $(L,M)$, namely the zero curvature equation given as
\begin{align}
	L_t -M_x + [L,M] = 0,
\end{align} 
generates a coupled system of variable-coefficient FLE with an additional term $\Gamma(t)$. The coupled system takes the form as
\begin{align}\label{cfle1}
	&q_{xt} - M(t)\ q+[2 i N(t)\ q\ r -\Gamma(t)]q_x   = 0\\ \label{cfle2}
	&r_{xt} - M(t)\ r -[2 i N(t)\ r\ q +\Gamma(t)]r_x   = 0.			
\end{align}
Here $q(x,t)$ and $r(x,t)$ are two different complex valued fields. The term $\Gamma(t)$ arises out of the Lax pair which takes the form as 
\begin{align}
	\Gamma(t)=\frac{W[N(t),M(t)]}{2\ N(t)\ M(t)},
\end{align} 
where the Wronskian is $W[N(t),M(t)]=N(t)M'(t)-M(t)N'(t)$. \\
\indent Under the local symmetry reduction $r(x,t) = q^*(x,t)$, the coupled Eqs.~(\ref{cfle1})-(\ref{cfle2}) yield the local variable-coefficient FLE, which has been studied for both zero and non-zero background in \citep{talukdar2026bright,dutta2025soliton}. However, when $M(t) = N(t) = \textit{1}$, it follows that $\Gamma(t) = 0$, and the system reduces to the standard FLE, as investigated by the authors in \citep{talukdar2023multi,dutta2023fokas}. \\
\indent Likewise, under the nonlocal symmetry reduction $r(x,t) = i\, q^*(-x,-t)$, the coupled system (\ref{cfle1})-(\ref{cfle2}) reduces to the $\mathcal{PT}$-invariant nonlocal variable-coefficient FLE given by Eqs. (\ref{cfle3})-(\ref{cfle4}). When $M(t) = N(t) = \textit{1}$, we again have $\Gamma(t) = 0$, and the system becomes the reverse space--time nonlocal FLE reported in \citep{zhang2022integrability}.\\

\section{Nonstandard Hirota bilinearization}
\label{section:3}
In this section, we intend to obtain the symmetry preserving and breaking solutions of system (\ref{cfle3})-(\ref{cfle4}) using the nonstandard Hirota bilinearization\citep{Hirota_2004}. To do so, we consider $q(x,t)$ and $q^*(-x,-t)$ as two independent fields. To bilinearize  Eqs. (\ref{cfle3})-(\ref{cfle4}), let us assume 
\begin{align}
	\label{bilin0}
	q(x,t) =\sqrt{\frac{M(t)}{N(t)}}\frac{g(x,t)}{f(x,t)}, \quad q^*(-x,-t)=\sqrt{\frac{M(t)}{N(t)}}\frac{g^*(-x,-t)}{f^*(-x,-t)},
\end{align}
where $g(x,t)$, $f(x,t)$, and $g^*(-x,-t)$, $f^*(-x,-t)$ are independent complex functions. Inserting Eq. (\ref{bilin0}) in  Eqs. (\ref{cfle3})-(\ref{cfle4}), we obtain a set of nonlinear  equations in terms of the newly introduced fields,
\begin{align}
	\label{Bilin1}
	& \frac{1}{f^2(x,t)} \left(D_x D_t - M(t)\right)\, g(x,t) \cdot f(x,t)
	- \frac{g(x,t)}{f^3(x,t)}\, D_x D_t \left(f(x,t) \cdot f(x,t)\right) \nonumber \\
	& - \frac{2 M(t)\, g(x,t)\, g^*(-x,-t)}{f^3(x,t)\, f^*(-x,-t)}\, D_x \left(g(x,t) \cdot f(x,t)\right) \nonumber \\
	& + \frac{s_1(x,t)\, g(x,t)\, g^*(-x,-t)}{f^3(x,t)} 
	- \frac{s_1(x,t)\, g(x,t)\, g^*(-x,-t)\, f^*(x,t)}{f^3(x,t)\, f^*(-x,-t)} = 0,
\end{align}

\begin{align}		
	\label{Bilin2}
	& \frac{1}{{{f^*}^2}(-x,-t)} \left(D_x D_t - M(t)\right)\, g^*(x,t) \cdot f^*(x,t)
	- \frac{g^*(x,t)}{{f^*}^3(x,t)}\, D_x D_t \left(f^*(x,t) \cdot f^*(x,t)\right) \nonumber \\
	& + \frac{2 M(t)\, g(x,t)\, g^*(-x,-t)}{{f^*}^3(-x,-t)\, f(x,t)}\, D_x \left(g^*(-x,-t) \cdot f^*(-x,-t)\right) \nonumber \\
	& + \frac{s_2(x,t)\, g^*(-x,-t)\, g(x,t)}{{f^*}^3(-x,-t)} 
	- \frac{s_2(x,t)\, g(x,t)\, g^*(-x,-t)\, f^*(-x,-t)}{f(x,t)\, {f^*}^3(-x,-t)} = 0.
\end{align} 
Here $D_x$ and $D_t $ are the  Hirota derivatives\citep{Hirota_2004}, which   are defined as,
\begin{align}
	D_x^m D_t^n g(x,t).f(x,t)= 
	(\frac{\partial}{\partial z}  - \frac{\partial}{\partial x^\prime})^m
	(\frac{\partial}{\partial t}  - \frac{\partial}{\partial t^\prime})^n
	g(x,t).f(x^\prime,t^\prime)\Bigg|_{ (x=x^\prime)(t= t^\prime)},
\end{align} 
where $m$ and $n$ are positive integers.

Notice that the last two terms in Eqs. (\ref{Bilin1})-(\ref{Bilin2}) contain two auxiliary functions $s_j(x,t)$, $j=1,2$ which are introduced so that Eqs. (\ref{Bilin1})-(\ref{Bilin2}) can be cast into the following bilinear equations,
\begin{align}
	\label{BR1}
	&\Big[D_x D_t -M(t)\Big](g(x,t).f(x,t))=0,\\
	\label{BR2}
	&D_x D_t (f(x,t).f(x,t))  = s_1(x,t)g^*(-x,-t),\\
	\label{BR3}
	&2 M(t) D_x (g(x,t).f(x,t))   = -s_1(x,t)f^*(-x,-t),
\end{align}
\begin{align}
	\label{BR4}
	&\Big[D_x D_t -M(t)\Big](g^*(-x,-t).f^*(-x,-t))=0,\\
	\label{BR5}
	&D_x D_t (f^*(-x,-t).f^*(-x,-t))  = s_2(x,t)g(x,t),\\
	\label{BR6}
	&2 M(t) D_x (g^*(-x,-t).f^*(-x,-t))   = s_2(x,t)f(x,t),
\end{align}
where $M(t)$ is the dispersion coefficient as discussed earlier.\\
\indent Subsequently, to obtain the soliton solutions, we express $g(x,t)$, $f(x,t)$, $s_1(x,t)$,  $g^*(-x,-t)$, $f^*(-x,-t)$, and $s_2(x,t)$, by suitable series representations with respect to an arbitrary parameter $\epsilon$ as follows:
\begin{align}
	\label{GF}
	g(x,t)&= \epsilon g_1 + \epsilon^3 g_3 +.....,\quad
	&&g^*(-x,-t)= \epsilon g_1^* + \epsilon^3 g_3^* +....., \\
	f(x,t)&= 1+ \epsilon^2 f_2 + \epsilon^4 f_4 +.....,\quad
	&&f^*(-x,-t)= 1 + \epsilon^2 f_2^* + \epsilon^4 f_4^* +....., \\
	s_1(x,t)&= \epsilon s_{11} + \epsilon^3 s_{13} + .....\quad , \label{GF2}
	&&s_2(x,t)= \epsilon s_{21} + \epsilon^3 s_{23}+......	
\end{align} 
\indent Therefore, substituting the above expansions  (\ref{GF})-(\ref{GF2}) into the bilinear equations given in Eqs. (\ref{BR1})-(\ref{BR6}), we end up with the set of linear PDEs for the unknown functions $g_1$, $g_1^*$, $f_2$, and $f_2^*$. After, systematically solving the sets of linear PDEs, with appropriate choice of seed solutions one can obtain the multi-soliton solutions.
\subsection{ One bright soliton solution and periodic waves}
To derive the general one soliton solution ({\color{blue}\textit{1}-SS}) of Eqs. (\ref{cfle3}) and (\ref{cfle4}), we consider $q(x,t)$ and $q^*(-x,-t)$ as two independent fields and the appropriate seed solutions are taken as $g_1=\alpha_1 e^{\theta_1}$, $g^*_1=\alpha^*_1 e^{\theta^*_1}$,  where $\theta_1(x,t)=p_1 x+\omega_1(t)$ and $\theta_1^*(-x,-t)=p_1^* x+\omega_1^*(t)$. \\
\indent After inserting the above seed solutions into the linear PDEs we find the unknown functions $g_1$, $f_2$, $g_1^*$, $f_2^*$, $s_{11}$, $s_{21}$ as
\begin{align}
	\label{G1}
	g_1&= \alpha_1  e^{\theta_1},  &&g_1^*= \alpha_1^*  e^{\theta_1^*}, \\
	\label{F1}
	f_2 &= \beta_1 e^{\theta_1 +  \theta_1^*},  &&f_2^*= \beta_1^* e^{\theta_1 +  \theta_1^*}, \\
	\label{S1}
	s_{11}&= c_1 e^{\theta_1},  &&s_{21}= c_2 e^{\theta_1^*}.
\end{align} 
Here, $\alpha_1$, $\beta_1$, $\alpha_1^*$, and $\beta_1^*$ are arbitrary distinct complex constants and $c_1=-2 p_1 \alpha_1 M(t)$, $c_2=2 p_1^* \alpha_1^* M(t)$. 
Therefore, explicitly we have,
\begin{align}
	\label{sol1}
	q(x,t)=\sqrt{\frac{M(t)}{N(t)}} \frac{\epsilon g_1}{1+ \epsilon^2 f_2}\Big|_{\epsilon=1},
\end{align}
\begin{align}
	\label{sol1c}
	q^*(-x,-t)=\sqrt{\frac{M(t)}{N(t)}} \frac{\epsilon g_1^*}{1 + \epsilon^2 f_2^*}\Big|_{\epsilon=1}.
\end{align}
The symbol $`*'$ in this paper does not denote the complex conjugate. So, the wave numbers $p_1$ and $p_1^*$ are two distinct quantities and not complex conjugate of each other, and so also with other parameters. This is a necessary condition for obtaining the symmetry preserving and breaking soliton solutions of Eqs. (\ref{cfle3})-(\ref{cfle4}).\\ 
\indent  The resultant bright {\color{blue}\textit{1}-SS} of the $\mathcal{PT}$-invariant nonlocal vcFLE is obtained as,
\begin{align}\label{solution}
	q(x,t)=\sqrt{\frac{M(t)}{N(t)}}\frac{\alpha_1e^{\theta_1}}{1+\beta_1 e^{\theta_1+\theta_1^*}},\\
	\label{solution2} q^*(-x,-t)=\sqrt{\frac{M(t)}{N(t)}}\frac{\alpha_1^*e^{\theta_1^*}}{1+\beta_1^* e^{\theta_1+\theta_1^*}}.
\end{align}
where,
\begin{align}\nonumber
	\omega_1(t)&= \int \frac{M(t)dt}{p_1}, \quad  \omega_1^*(t)= \int \frac{M(t)dt}{p_1^*}, \\ \nonumber
	\beta_1 &=  -\frac{\alpha_1 \alpha_1^* p_1^2 p_1^* }{ (p_1+p_1^*)^2 }, \quad \beta_1^*=  \frac{\alpha_1 \alpha_1^* {p_1^*}^2 p_1}{ (p_1+p_1^*)^2}.
\end{align}
\indent It is a straightforward task to see the correctness of the solutions (\ref{solution}) and (\ref{solution2}) by putting it back into Eqs. (\ref{cfle3})-(\ref{cfle4}). It is important to note that solutions (\ref{solution}) and (\ref{solution2}) are not the complex conjugate of each other i.e. they are independent of each other. \\
Now, we write the solutions (\ref{solution}) and (\ref{solution2}) in terms of hyperbolic functions. Explicitly, we have
\begin{align}\label{sol46}
	q(x,t)=\sqrt{\frac{M(t)}{N(t)}} \frac{\alpha_1}{2} sech[\frac{1}{2}(\theta_1+\theta_1^*+\ln \beta_1)]\ e^{\frac{1}{2}(\theta_1-\theta_1^*-\ln \beta_1)},\\ 
	\label{sol47}
	q^*(-x,-t)=\sqrt{\frac{M(t)}{N(t)}} \frac{\alpha_1^*}{2} sech[\frac{1}{2}(\theta_1^*+\theta_1+\ln \beta_1^*)]\ e^{\frac{1}{2}(\theta_1^*-\theta_1-\ln \beta_1^*)},
\end{align} 
where the parameters $\theta_1$, $\theta_1^*$, $\alpha_1$, $\alpha_1^*$, $\beta_1$, and $\beta_1^*$ have already been introduced earlier. 
Now, let us rewrite the solutions (\ref{solution}) and (\ref{solution2}) by considering $p_1=a+ib$ and $p_1*=m+in$, obtaining the explicit form as,
\begin{align}\label{1ss1}
	q(x,t)=\sqrt{\frac{M(t)}{N(t)}}\frac{\text{$\alpha_1$}\ e^{\eta +i \chi } }{\text{1}-\frac{ \text{$\alpha_1 $} \text{$\alpha_1$}^* (a+i b)^2 (m+i n)}{ (a+i b+m+i n)^2}e^{\left(\eta ^*+\eta \right)+i \left(\chi ^*+\chi \right)}},\\ \label{1ss2} 
	q^*(-x,-t)=\sqrt{\frac{M(t)}{N(t)}}\frac{\text{$\alpha_1$}^* e^{\eta ^*+i \chi ^*} }{\text{1}+\frac{\text{$\alpha_1$} \text{$\alpha_1$}^* (a+i b) (m+i n)^2}{ (a+i b+m+i n)^2}e^{\left(\eta ^*+\eta \right)+i \left(\chi ^*+\chi \right)}}.
\end{align}
Here, the variables are $\eta=a\left(x+ \frac{ \int M(t)dt }{a^2+b^2}\right)$, $\chi=b\left( x-\frac{ \int M(t)dt }{a^2+b^2}\right) $ \& $\eta^*=m\left(x+\frac{ \int M(t)dt }{m^2+n^2}\right)$, $\chi^*=n \left(x-\frac{ \int M(t)dt }{m^2+n^2}\right)$ for the complex fields $q(x,t)$ and $q^*(-x,-t)$ respectively. The velocities are $c(t) =\frac{ \int M(t)dt }{a^2+b^2}$ and $c^*(t) = \frac{ \int M(t)dt }{m^2+n^2}$ for (\ref{1ss1}) and (\ref{1ss2}), respectively.\\
\indent The set of solutions presented in (\ref{1ss1})-(\ref{1ss2}) constitutes the most general one bright soliton solution to Eqs. (\ref{cfle3})-(\ref{cfle4}). We refer this as a symmetry-broken solution, as the fields  are generally independent and cannot be obtained from one another i.e. under the joint action of the $\mathcal{PT}$-operator solutions (\ref{1ss1}) and (\ref{1ss2}) are not interreducible to one another.\\ 
\indent In order to obtain the symmetry preserving solution for the Eqs. (\ref{cfle3})-(\ref{cfle4}) we consider $\alpha_1=\alpha_1^*=1$. Additionally, we impose a restriction on the wave numbers $p_1$ \& $p_1^*$ such that under the condition $a=m=0$ and $p_1^*=p_1$, it leads to the relation $b=n$. Under this condition on the wave numbers, the solutions given by (\ref{1ss1})-(\ref{1ss2}) preserve the $\mathcal{PT}$ symmetry. We obtain the $\mathcal{PT}$-preserving form as, \\
\begin{align}\label{ptsol1}
	q(x,t)=\sqrt{\frac{M(t)}{N(t)}}\frac{e^{i \chi } }{1-\frac{ i b }{4}e^{2i\chi}},\\
	\label{ptsol2}
	q^*(-x,-t)=\sqrt{\frac{M(t)}{N(t)}}\frac{e^{i \chi ^*} }{1+\frac{ i n}{4}e^{2i\chi ^*}},
\end{align} 
where $\chi=b\left(x- c(t)\right) $, $\chi^*=n \left(x-c^*(t)\right)$ and the two velocities are  $c(t)=\frac{ \int M(t)dt }{b^2}$, $c^*(t)=\frac{ \int M(t)dt }{n^2}$, respectively. One can easily see that the two solutions given by (\ref{ptsol1}) and (\ref{ptsol2}) reduce to one another under the action of space, time reversal and complex conjugation, i.e. 
\begin{align}
	q^*(-x,-t)=[q(x,t)]^*_{|{x\rightarrow-x; \ t\rightarrow-t}},
\end{align}  
and vice-versa, provided the dispersion \& nolinearity coefficients $M(t)$, $N(t)$ are even functions in time, $c(t)=c^*(t)$ and $\Gamma(t)$ are odd functions, which are necessary conditions for the system given by Eqs. (\ref{cfle3})-(\ref{cfle4}) to preserve the symmetry as discussed in Sec. \ref{section:2}. 
\begin{figure}[h!]
	\centering
	\captionsetup[subfigure]{skip=2pt}  
	
	\begin{subfigure}{0.30\textwidth}
		\centering
		\includegraphics[width=\textwidth]{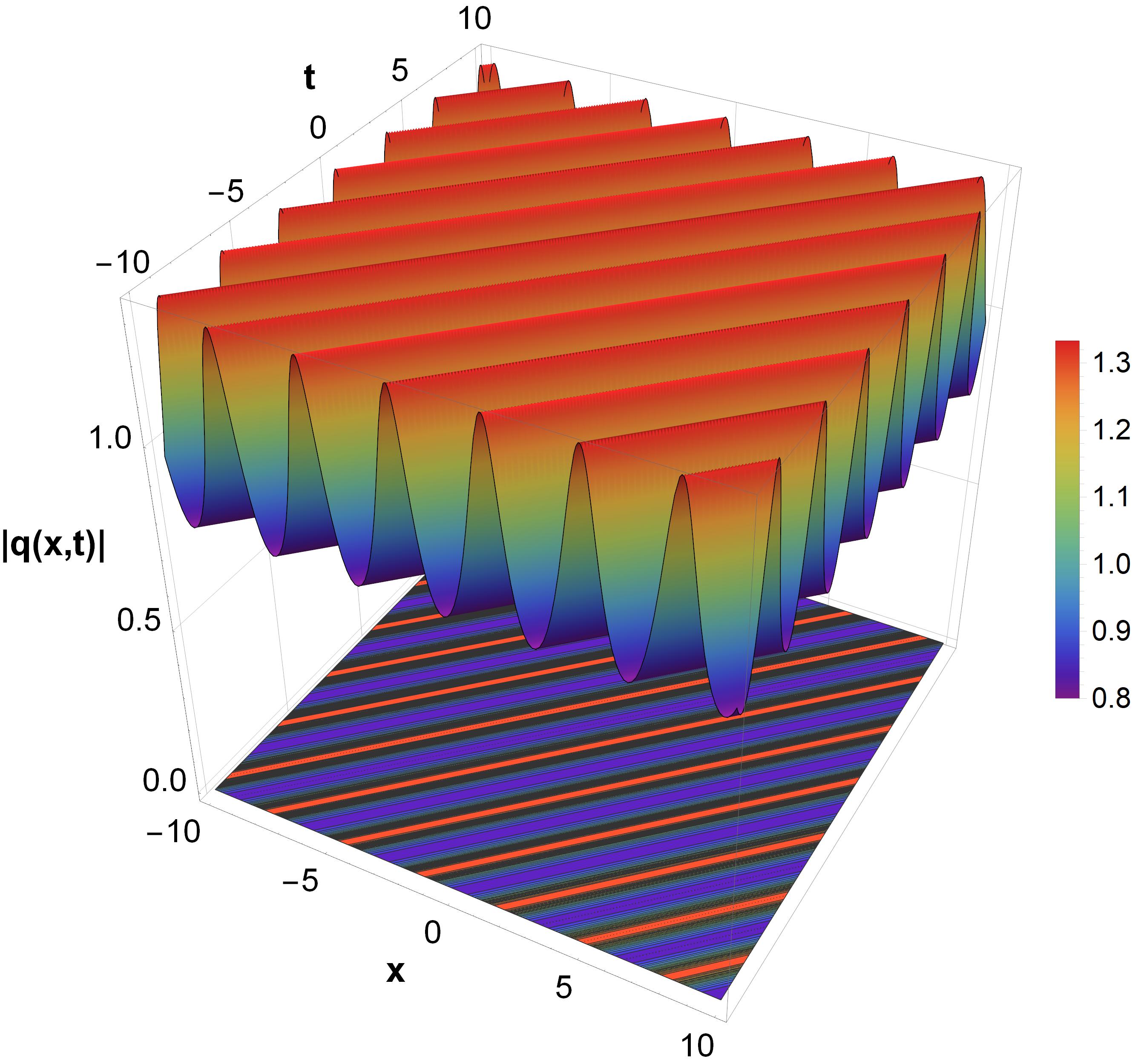}
		\caption{}
		\label{fig:1a}
	\end{subfigure}\hfill
	\begin{subfigure}{0.30\textwidth}
		\centering
		\includegraphics[width=\textwidth]{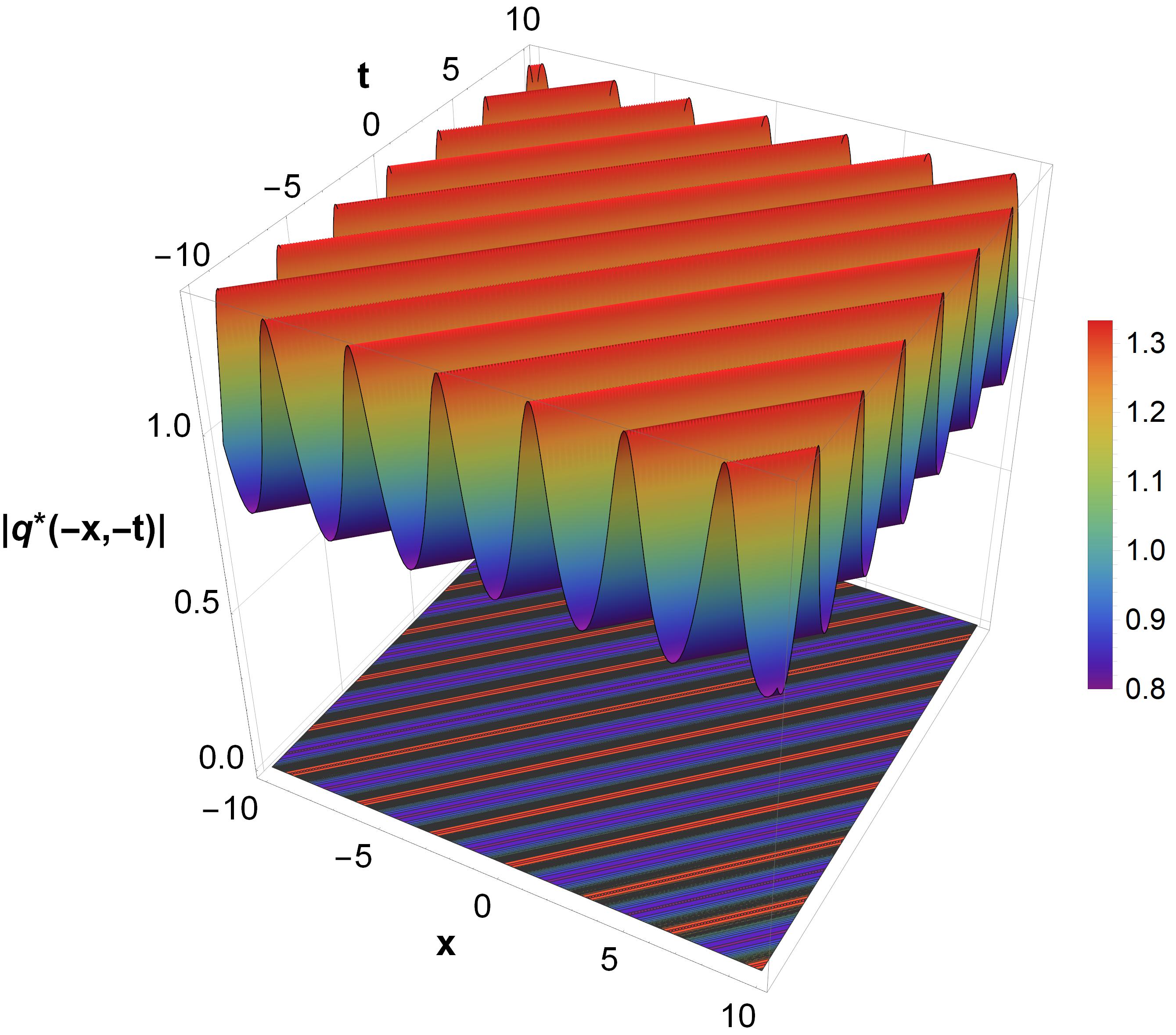}
		\caption{}
		\label{fig:1b}
	\end{subfigure}\hfill
	\begin{subfigure}{0.30\textwidth}
		\centering
		\includegraphics[width=\textwidth]{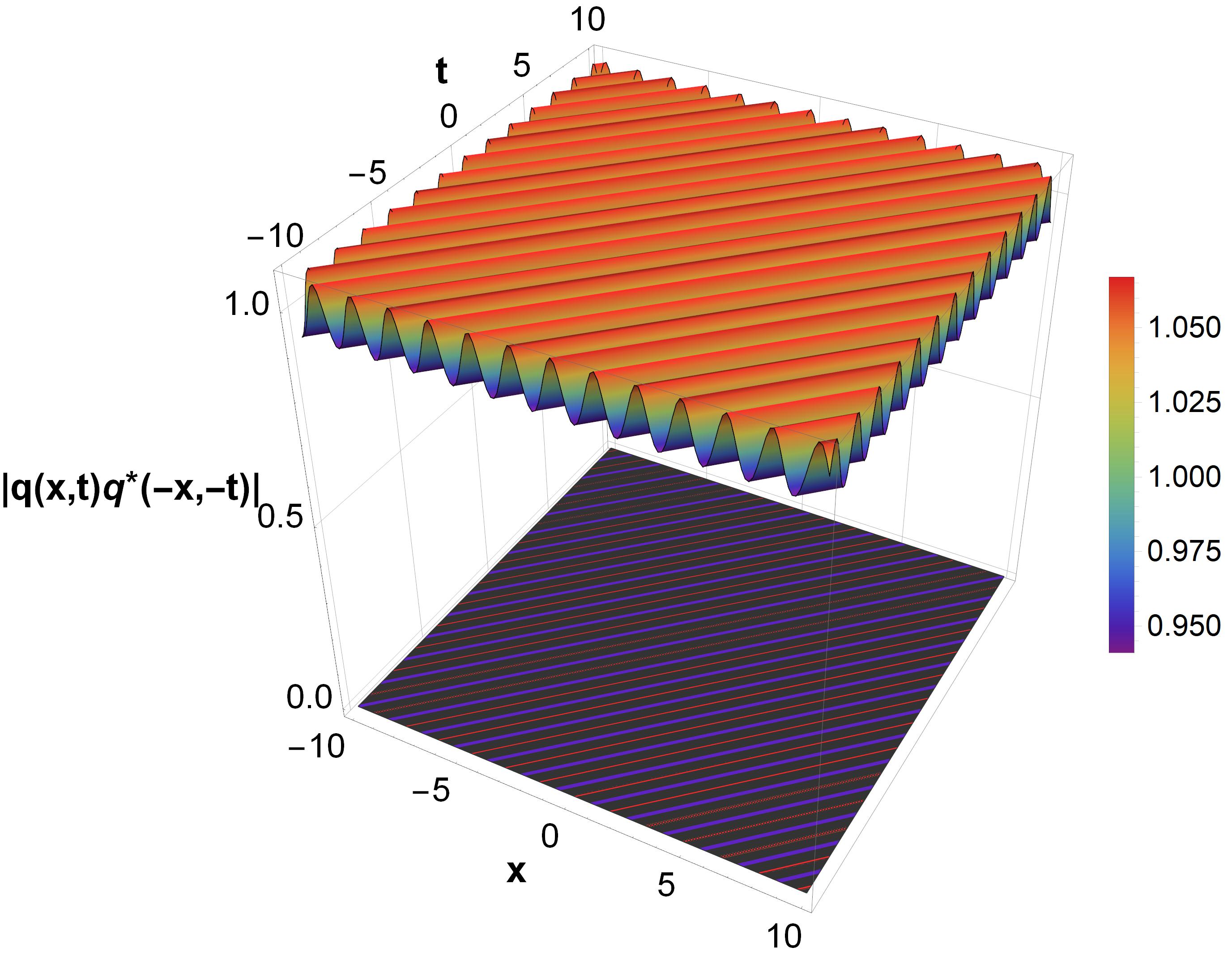}
		\caption{}
		\label{fig:1c}
	\end{subfigure}
	
	\vspace{0.3em}
	
	\begin{subfigure}{0.30\textwidth}
		\centering
		\includegraphics[width=\textwidth]{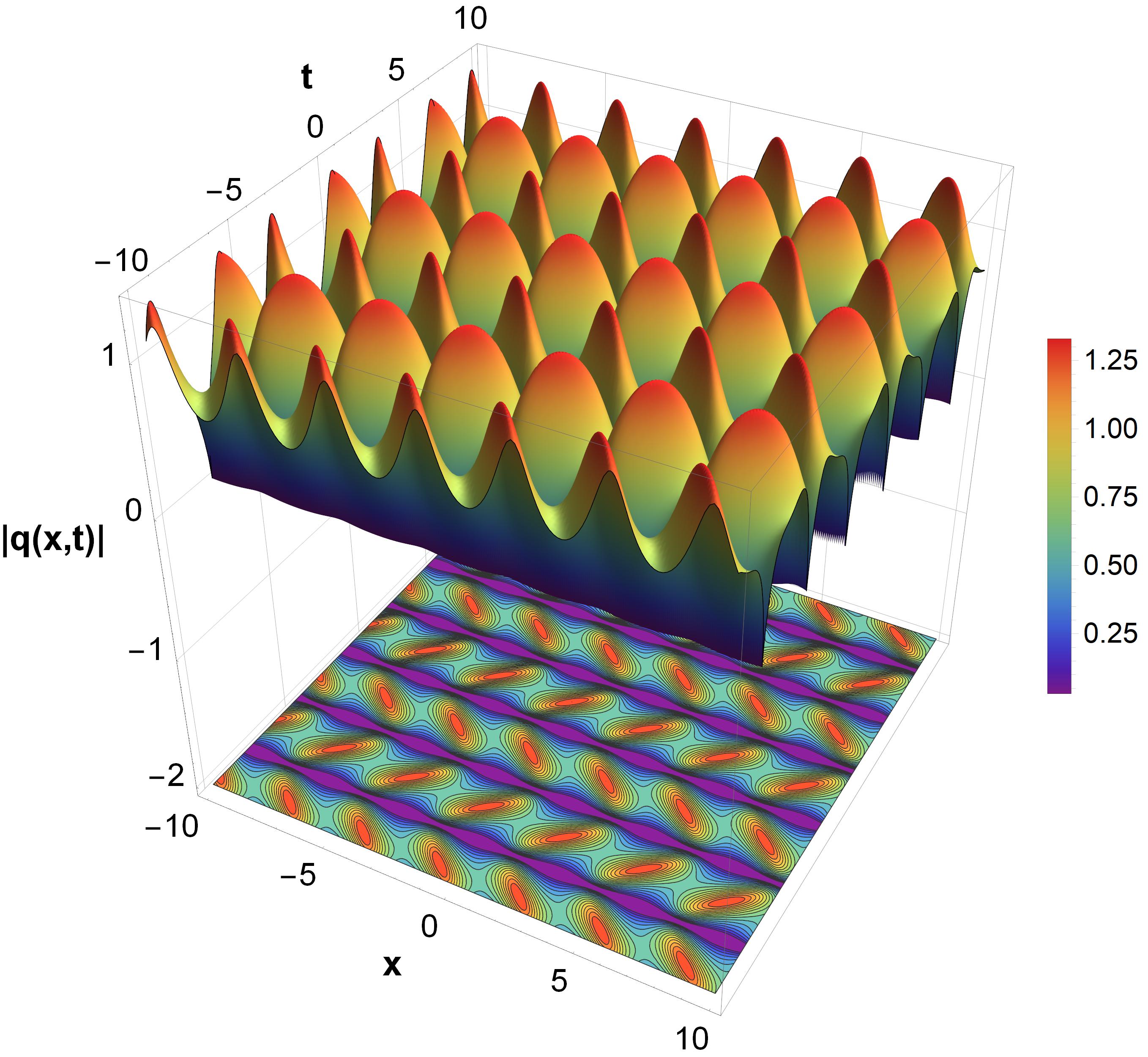}
		\caption{}
		\label{fig:1d}
	\end{subfigure}\hfill
	\begin{subfigure}{0.30\textwidth}
		\centering
		\includegraphics[width=\textwidth]{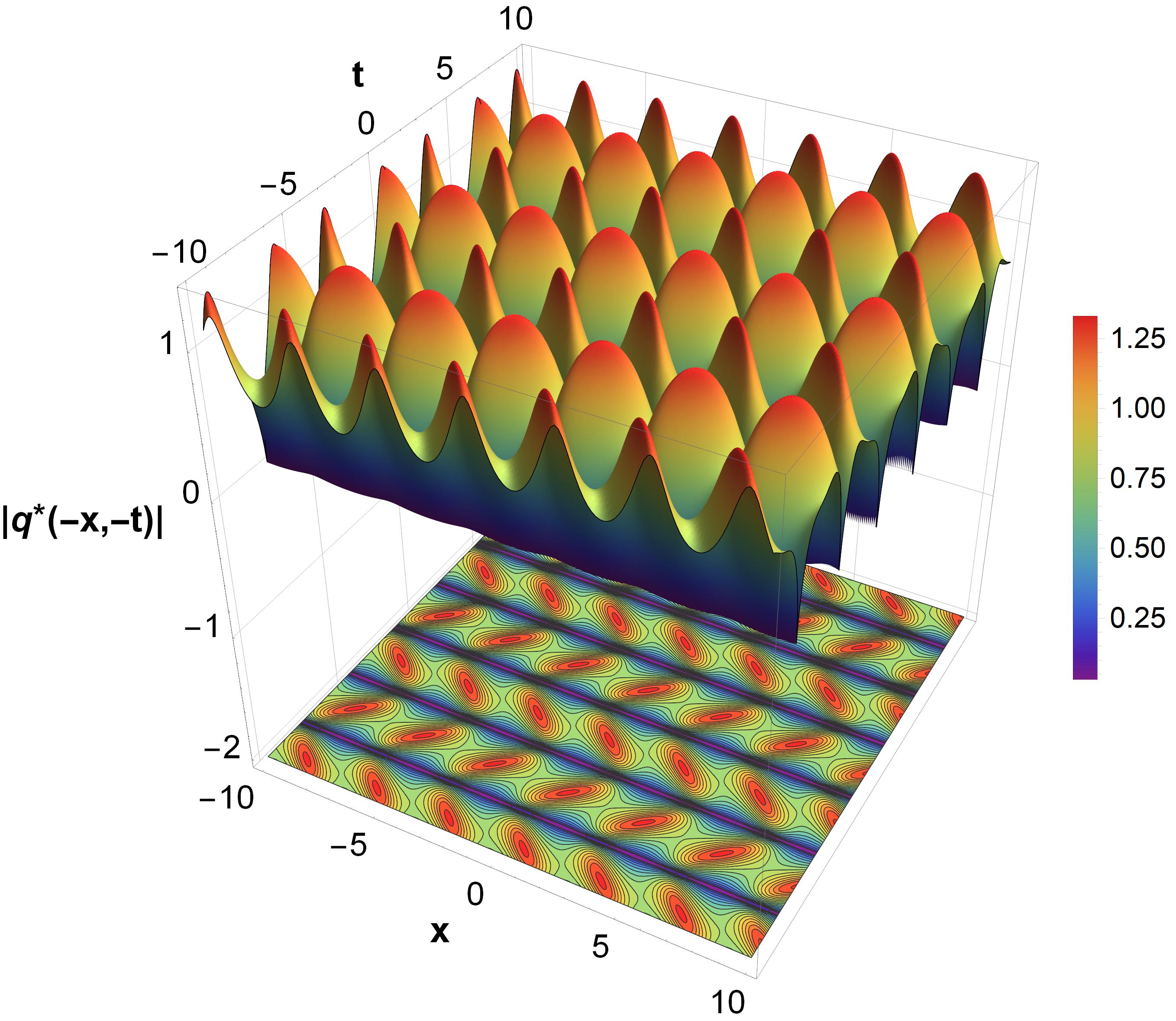}
		\caption{}
		\label{fig:1e}
	\end{subfigure}\hfill
	\begin{subfigure}{0.30\textwidth}
		\centering
		\includegraphics[width=\textwidth]{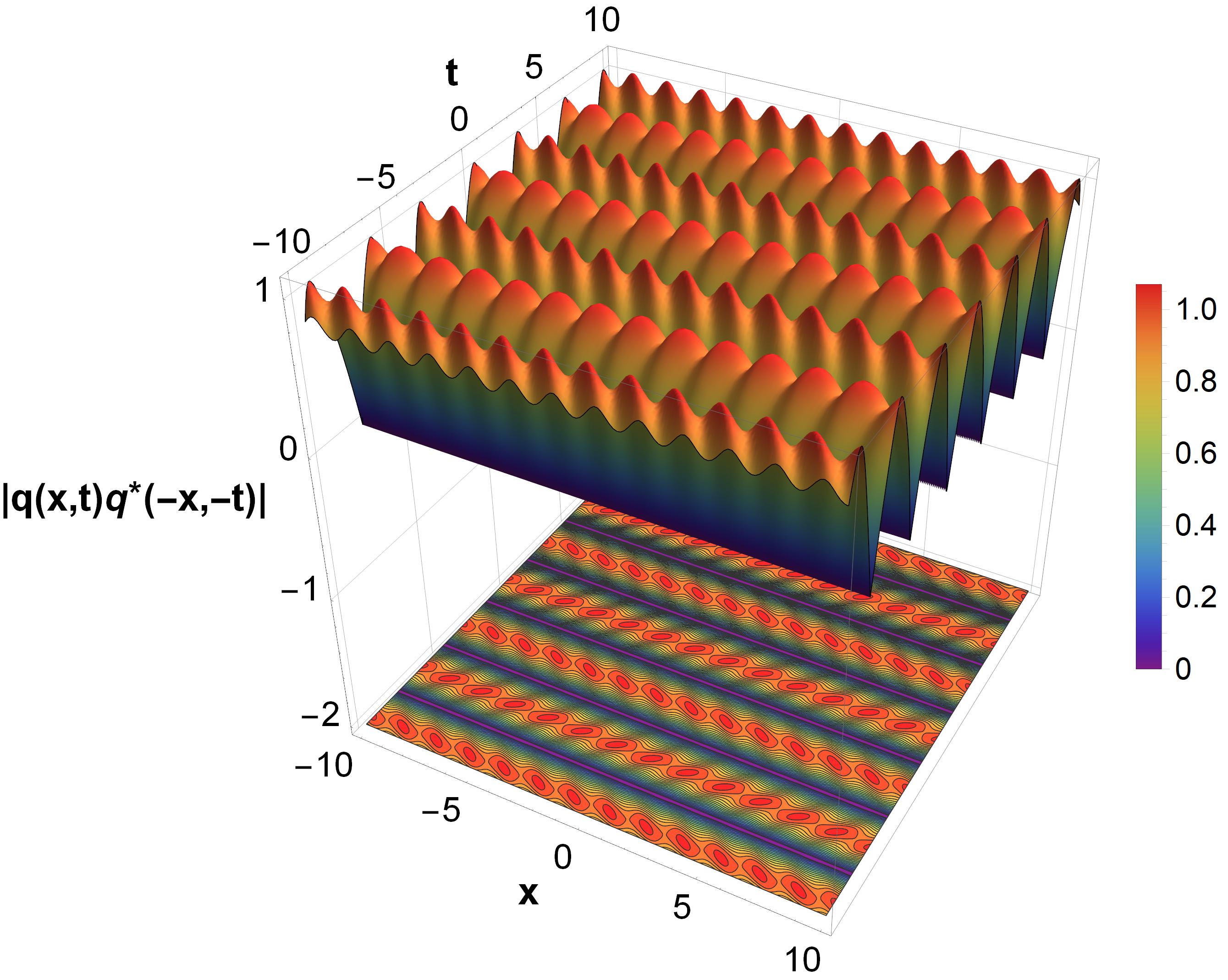}
		\caption{}
		\label{fig:1f}
	\end{subfigure}
	
	\vspace{0.3em}
	
	\begin{subfigure}{0.30\textwidth}
		\centering
		\includegraphics[width=\textwidth]{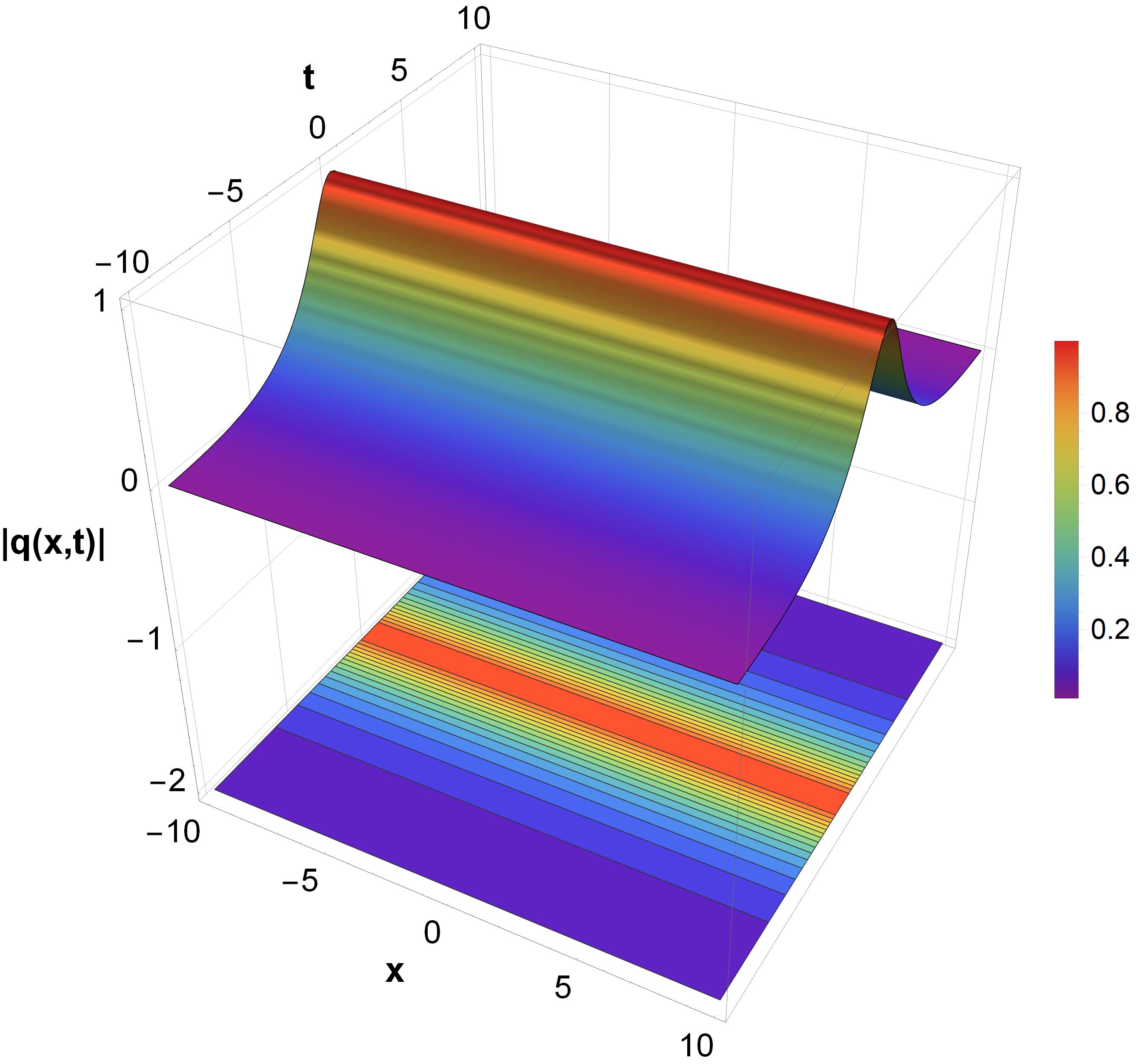}
		\caption{}
		\label{fig:1g}
	\end{subfigure}\hfill
	\begin{subfigure}{0.30\textwidth}
		\centering
		\includegraphics[width=\textwidth]{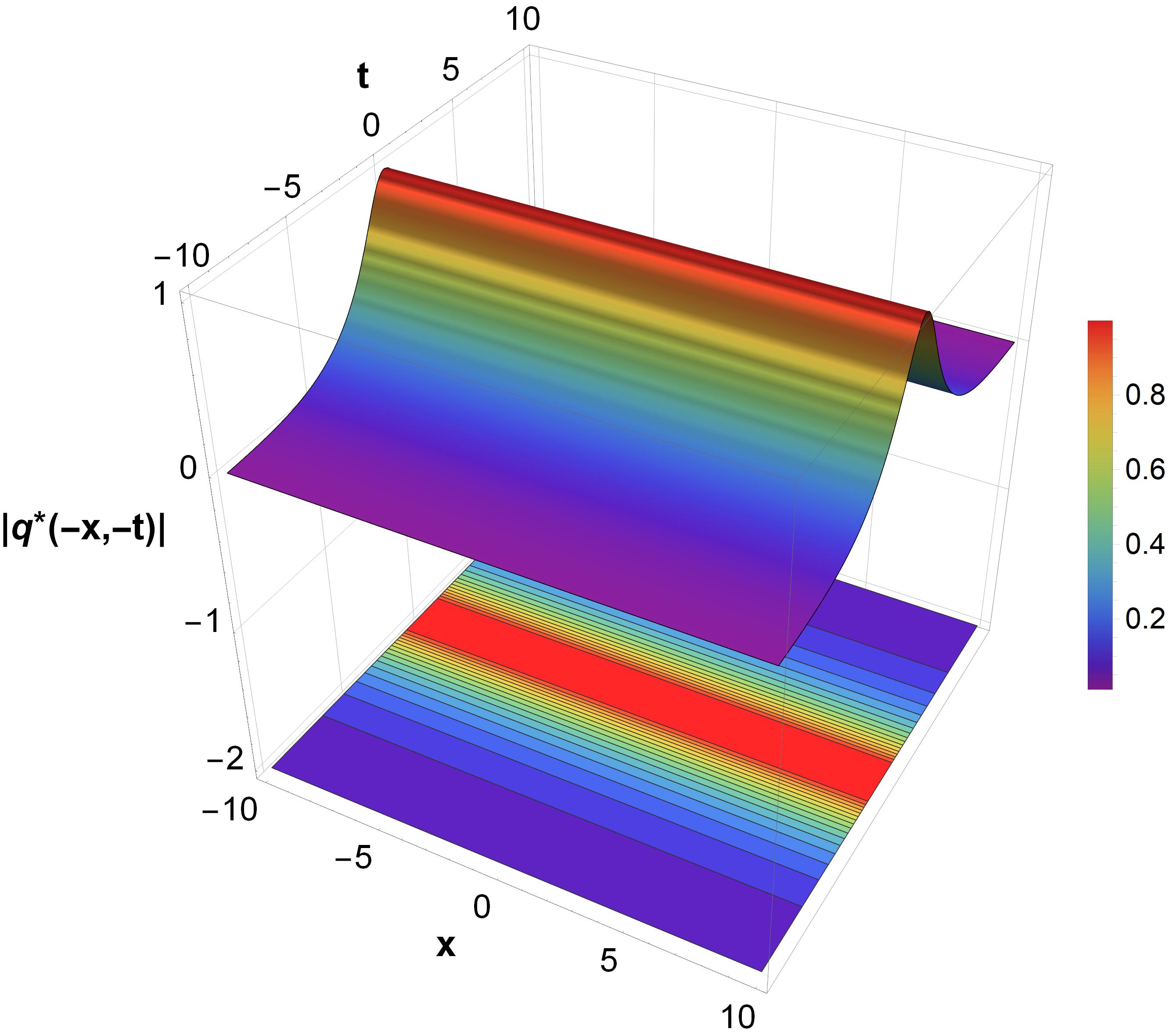}
		\caption{}
		\label{fig:1h}
	\end{subfigure}\hfill
	\begin{subfigure}{0.30\textwidth}
		\centering
		\includegraphics[width=\textwidth]{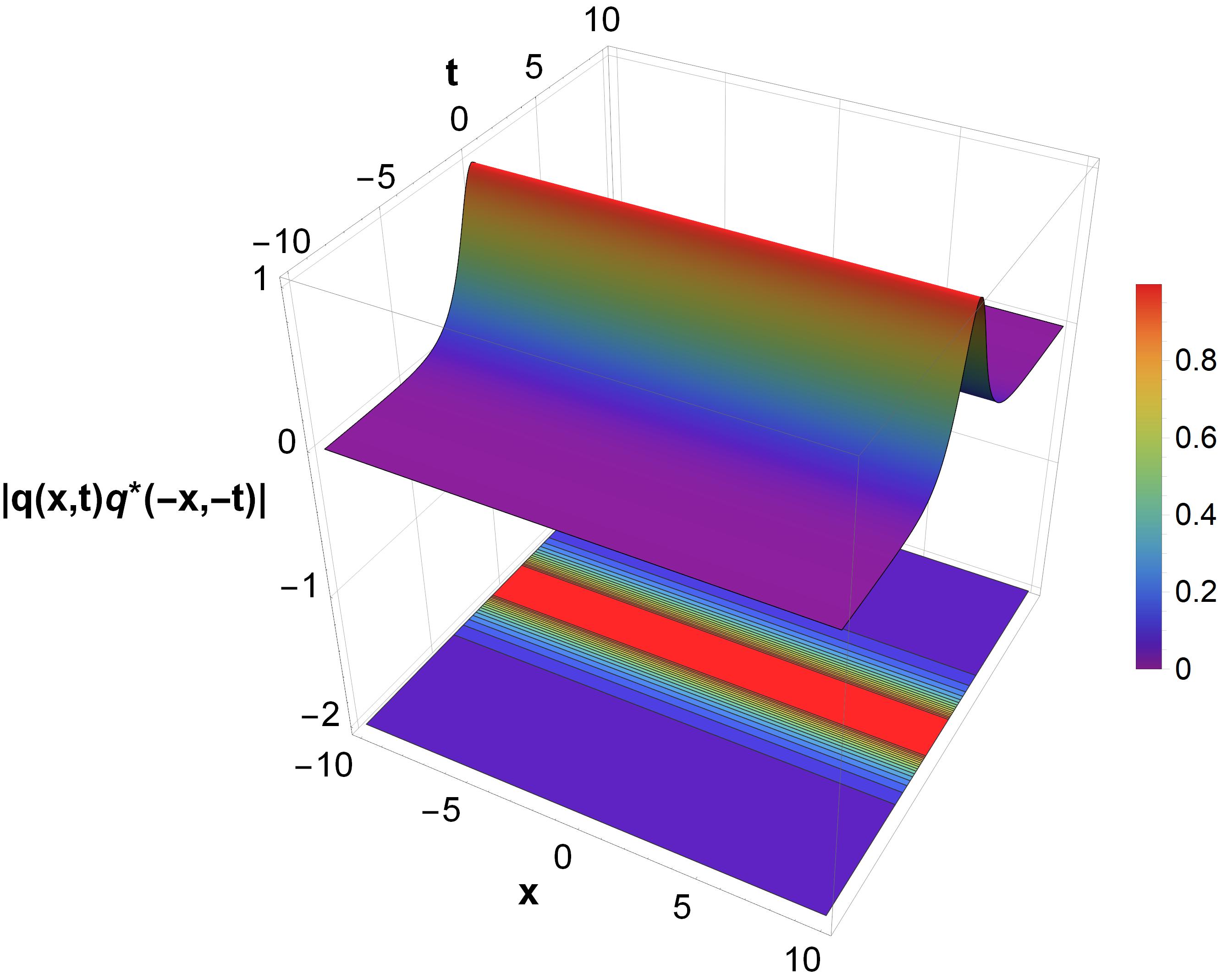}
		\caption{}
		\label{fig:1i}
	\end{subfigure}
	
	\vspace{0.3em}
	
	\caption{\scriptsize The plot in (a), (b) and (c) are periodic waves which 
		appears when $M(t)=1$, $N(t)=1$, $\Gamma(t)=0$ with $p_1=i$ and $p_1^*=i$ 
		for $|q(x,t)|$, $|q^*(-x,-t)|$ and $|q(x,t)q^*(-x,-t)|$, respectively. 
		In (d), (e) and (f) we obtain again a periodic wave structure when 
		$M(t)=\cos t$, $N(t)=1$, $\Gamma(t)=-\frac{1}{2}\tan t$ with $p_1=i$ 
		and $p_1^*=i$ for envelopes $|q(x,t)|$, $|q^*(-x,-t)|$ and 
		$|q(x,t)q^*(-x,-t)|$, respectively. In (g), (h) and (i) we obtain soliton 
		structure when $M(t)=\text{sech}\, t$, $N(t)=1$, 
		$\Gamma(t)=-\frac{1}{2}\tanh t$ with $p_1=0.001i$ and $p_1^*=0.001i$ 
		in $|q(x,t)|$, $|q^*(-x,-t)|$ and $|q(x,t)q^*(-x,-t)|$, respectively. 
		The other parametric constants are taken to be $\alpha_1=\alpha_1^*=1$.}
	\label{fig:one-solitons}
\end{figure}
Now, we express the solutions (\ref{ptsol1}) and (\ref{ptsol2}) explicitly in terms of the intensity profile as,
\begin{align}\label{intensity}
	q(x,t)q^*(-x,-t)=\frac{2M(t)}{bN(t)}\ sech\left(2i\chi+\ln\frac{b}{4}\right).
\end{align}
In the intensity profile given in (\ref{intensity}), the hyperbolic secant function carries an imaginary component, which makes it difficult to directly generate the corresponding intensity plot. To clarify the behaviour of the solution, we therefore also include the plots of $|q(x,t) q^{*}(-x,-t)|$ along with $|q(x,t)|$ and $|q^*(-x,-t)|$.\\ 
\indent In addition, we examine how a time-dependent dispersion profile influences the periodic waves obtained under constant dispersion and nonlinearity. To carry out this study, we allow $M(t)$ to take specific trigonometric and hyperbolic forms, ensuring that it remains an even function similar to $N(t)$, while $\Gamma(t)$ and $c(t)=c^{*}(t)$ are chosen as odd functions. To satisfy these criteria, we first consider $M(t)=cos t$, and in a second example, we take $M(t)=\mathrm{sech}\, t$. The chosen form of $M(t)$ directly affects the periodic structures as shown in Figs.~\ref{fig:1a}--\ref{fig:1c}, modifying their shape and orientation. For instance, selecting $M(t)=cos t$, which is itself periodic, naturally produces periodic wave patterns. Thus, when the dispersion and nonlinearity coefficients are taken as $M(t)=cos t$ and $N(t)=1$, the $\mathcal{PT}$-invariant nonlocal vcFLE shows periodic wave solutions, illustrated in Figs.~\ref{fig:1d}--\ref{fig:1f} for $|q(x,t)|$, $|q^{*}(-x,-t)|$, and $|q(x,t) q^{*}(-x,-t)|$, respectively. In this case, the imposed periodic dispersion $M(t)=cos t$ causes a noticeable change in the orientation of the wave pattern, as apparent from Figs.~\ref{fig:1d}--\ref{fig:1f}.\\
\indent Similarly, when the dispersion and nonlinearity coefficients are chosen to be $M(t)=sech\ t$ and $N(t)=1$, the periodic waves in Figs. \ref{fig:1a}-\ref{fig:1c} change and form soliton structures as illustrated in Figs. \ref{fig:1g}-\ref{fig:1i} in  $|q(x,t)|$, $|q^{*}(-x,-t)|$, and $|q(x,t) q^{*}(-x,-t)|$ accordingly. It is worth noting that only under the choice $M(t)=sech\ t$ we observe soliton formation in Eqs. (\ref{cfle3})-(\ref{cfle4}); otherwise the $\mathcal{PT}$-invariant nonlocal vcFLE yields periodic waves in the symmetry preserving condition.  

\subsection{ Bright two soliton solution and periodic waves}   

The two soliton solution ({\color{blue}\textit{2}-SS}) of Eqs. (\ref{cfle3})-(\ref{cfle4}) is obtained when we consider the seed solutions as $g_1=\alpha_1 e^{\theta_1}+\alpha_2 e^{\theta_2}$, $g^*_1=\alpha^*_1 e^{\theta^*_1}+\alpha_2^* e^{\theta_2^*}$,  where $\theta_1(x,t)=p_1 x+\omega_1(t)$, $\theta_1^*(-x,-t)=p_1^* x+\omega_1^*(t)$ and $\theta_2(x,t)=p_2 x+\omega_2(t)$ and $\theta_2^*(-x,-t)=p_2^* x+\omega_2^*(t)$. Inserting the above seed solutions in the linear PDEs, we find the unknown functions $f_2$, $g_3$, $f_4$, $f_2^*$, $g_3^*$, $f_4^*$ as 
\begin{align}
	\label{G2}
	f_2&= \beta_1 e^{\theta_1+\theta^*_1}  +  
	\beta_2 e^{\theta_2+\theta^*_2}+
	\beta_3 e^{\theta_1+\theta^*_2} + 
	\beta_4 e^{\theta_2+\theta^*_1},\\
	f_2^*&= \beta_1^* e^{\theta_1+\theta^*_1}  +
	\beta_2^* e^{\theta_2+\theta^*_2}+  
	\beta_3^* e^{\theta^*_1+\theta_2} + 
	\beta_4^* e^{\theta^*_2+\theta_1} ,\\
	g_3 &= \alpha_3 e^{\theta_1+\theta^*_1+\theta_2} 
	+\alpha_4 e^{\theta_1+\theta_2+\theta^*_2}, \\
	g_3^* &= \alpha_3^* e^{\theta_1+\theta^*_1+\theta_2^*} 
	+\alpha_4^* e^{\theta_1^*+\theta_2+\theta^*_2 }, \\
	f_4 & =\beta_5
	e^{\theta_1+\theta^*_1+\theta_2+\theta^*_2}, \\
	f_4^* & =\beta_5^*
	e^{\theta_1+\theta^*_1+\theta_2+\theta^*_2} .
\end{align}
The auxiliary functions are written as 	$s_1(x,t)= c_{11} e^{\theta_1}+ c_{12} e^{\theta_2} + c_{13} e^{\theta_1+\theta^*_1+\theta_2} + c_{14} e^{ \theta_1+\theta_2+\theta^*_2}$ and $	s_2(x,t)= c_{21} e^{\theta_1}+ c_{22} e^{\theta_2} + c_{23} e^{\theta_1+\theta^*_1+\theta_2} + c_{24} e^{ \theta_1+\theta_2+\theta^*_2}$. Explicitly, we write    
\begin{align}
	\label{2sol}
	q(x,t)=\sqrt{\frac{M(t)}{N(t)}}\frac{\epsilon g_1 + \epsilon^3 g_3}{ 1 + \epsilon^2 f_2 + \epsilon^4 f_4}\Big|_{\epsilon=1} ,
\end{align} 
\begin{align}
	\label{2solc}
	q^*(-x,-t)=\sqrt{\frac{M(t)}{N(t)}}\frac{\epsilon g_1^* + \epsilon^3 g_3^*}{ 1 + \epsilon^2 f_2^* + \epsilon^4 f_4^*}\Big|_{\epsilon=1}. 
\end{align}
Now, substituting the unknown functions in Eqs. (\ref{2sol})-(\ref{2solc}) we obtain the following explicit form of {\color{blue}\textit{2}-SS}:
\begin{align}
	\label{sol3}
	q(x,t)=\sqrt{\frac{M(t)}{N(t)}} \frac{\alpha_1 e^{\theta_1}+\alpha_2 e^{\theta_2}+\alpha_3 e^{\theta_1+\theta_1^*+\theta_2}+\alpha_4 e^{\theta_2+\theta_2^*+\theta_1}}{1 + \beta_1 e^{\theta_1+\theta_1^*}+ \beta_2 e^{\theta_2+\theta_2^*}+ \beta_3 e^{\theta_1+\theta_2^*}+ \beta_4 e^{\theta_1^*+\theta_2}+ \beta_5 e^{\theta_1+\theta_1^*+\theta_2+\theta_2^*}},\\
	\label{sol4}
	q(-x,-t)=\sqrt{\frac{M(t)}{N(t)}} \frac{\alpha_1^* e^{\theta_1^*}+\alpha_2^* e^{\theta_2^*}+\alpha_3^* e^{\theta_1^*+\theta_1+\theta_2^*}+\alpha_4^* e^{\theta_2^*+\theta_2+\theta_1^*}}{1 + \beta_1^* e^{\theta_1+\theta_1^*}+ \beta_2^* e^{\theta_2+\theta_2^*}+ \beta_3^* e^{\theta_1^*+\theta_2}+ \beta_4^* e^{\theta_1+\theta_2^*}+ \beta_5^* e^{\theta_1+\theta_1^*+\theta_2+\theta_2^*}},
\end{align}
where
\begin{align*}
	\omega_1(t)&= \int \frac{M(t)dt}{p_1}, \quad  \omega_1^*(t)= \int \frac{M(t)dt}{p_1^*}, \quad
	\omega_2(t)=\int \frac{M(t)dt}{p_2}, \quad  \omega_2^*(t)=\int \frac{M(t)dt}{p_2^*}, \\
	\alpha_3 &=\frac{ \alpha_1 \alpha_1^* \alpha_2 {p^*_1}^3 (p_1-p_2)^2}{ (p_1-p^*_1)^2 (p_2-p^*_1)^2},
	\quad
	\alpha_3^*=\frac{(p_1^* - p_2^*)^2 \, p_1^3 \, \alpha_1 \, \alpha_1^*\, \alpha_2^*}{(p_2^* + p_1)^2 \, (p_1^* + p_1)^2 },\\
	\alpha_4 &=
	- \frac{(p_1 - p_2)^2 \, {p_2^*}^3 \, \alpha_2 \, \alpha_2^* \, \alpha_1}{(p_1 + p_2^*)^2 \, (p_2 + p_2^*)^2 },\quad
	\alpha_4^* = \frac{(p_1^* - p_2^*)^2 \, p_2^3 \, \alpha_2 \, \alpha_2^* \, \alpha_1^*}{(p_1^* + p_2)^2 \, (p_2^* + p_2)^2 },\\
	\end{align*}
    \begin{figure}[]
	\centering
	\captionsetup[subfigure]{skip=2pt}
	
	\begin{subfigure}{0.30\textwidth}
		\centering
		\includegraphics[width=\textwidth]{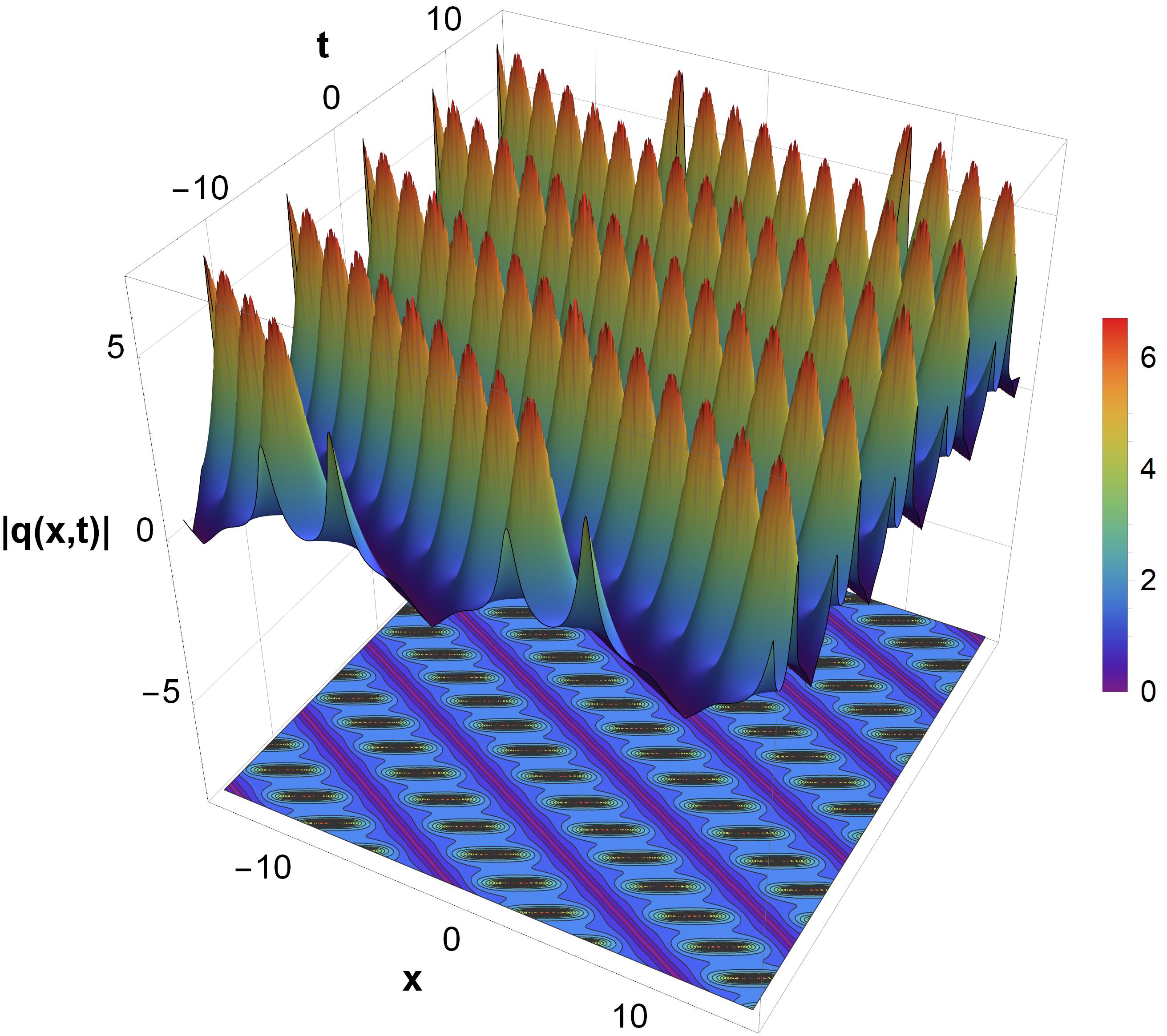}
		\caption{}
		\label{fig:2a}
	\end{subfigure}\hfill
	\begin{subfigure}{0.30\textwidth}
		\centering
		\includegraphics[width=\textwidth]{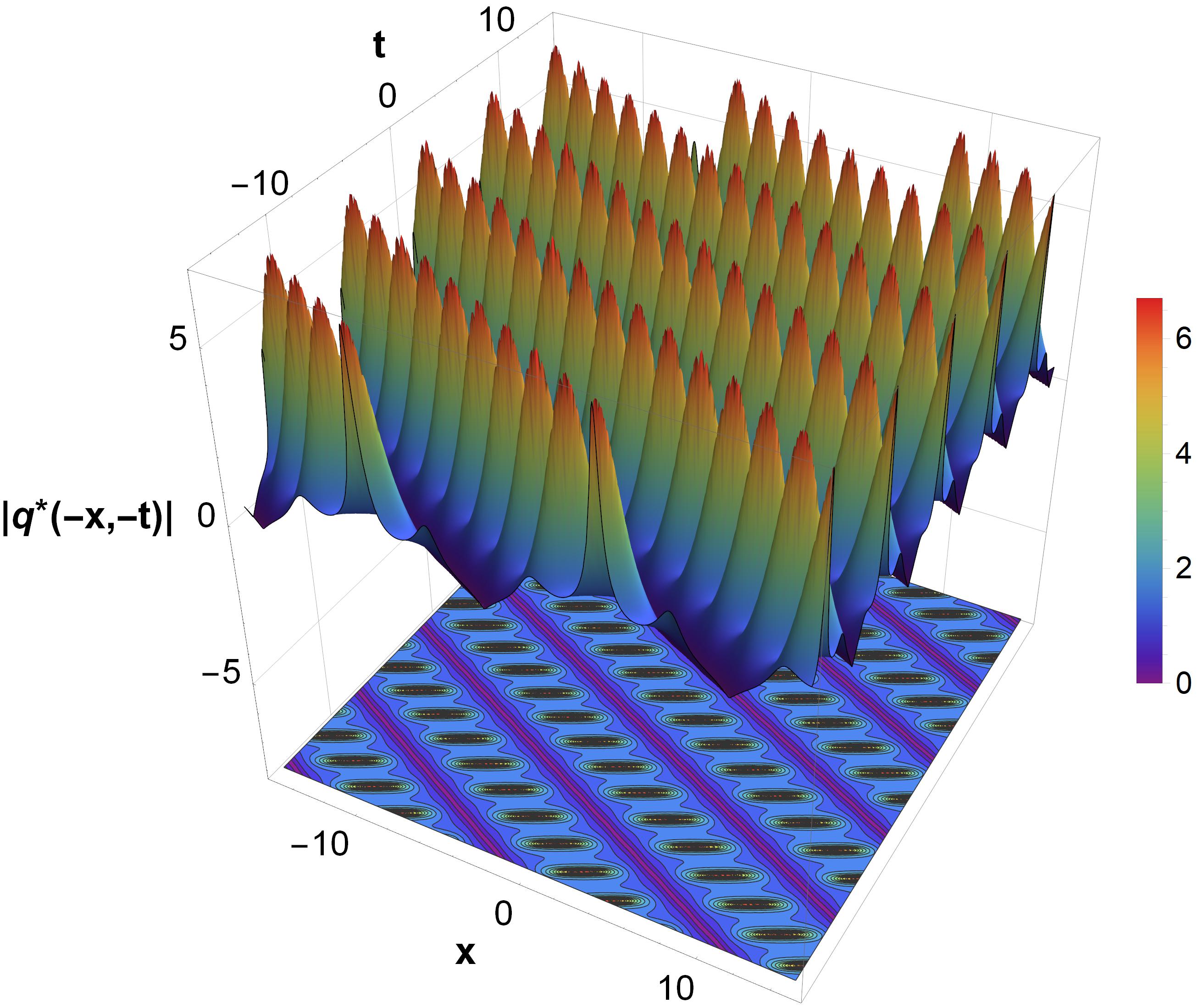}
		\caption{}
		\label{fig:2b}
	\end{subfigure}\hfill
	\begin{subfigure}{0.30\textwidth}
		\centering
		\includegraphics[width=\textwidth]{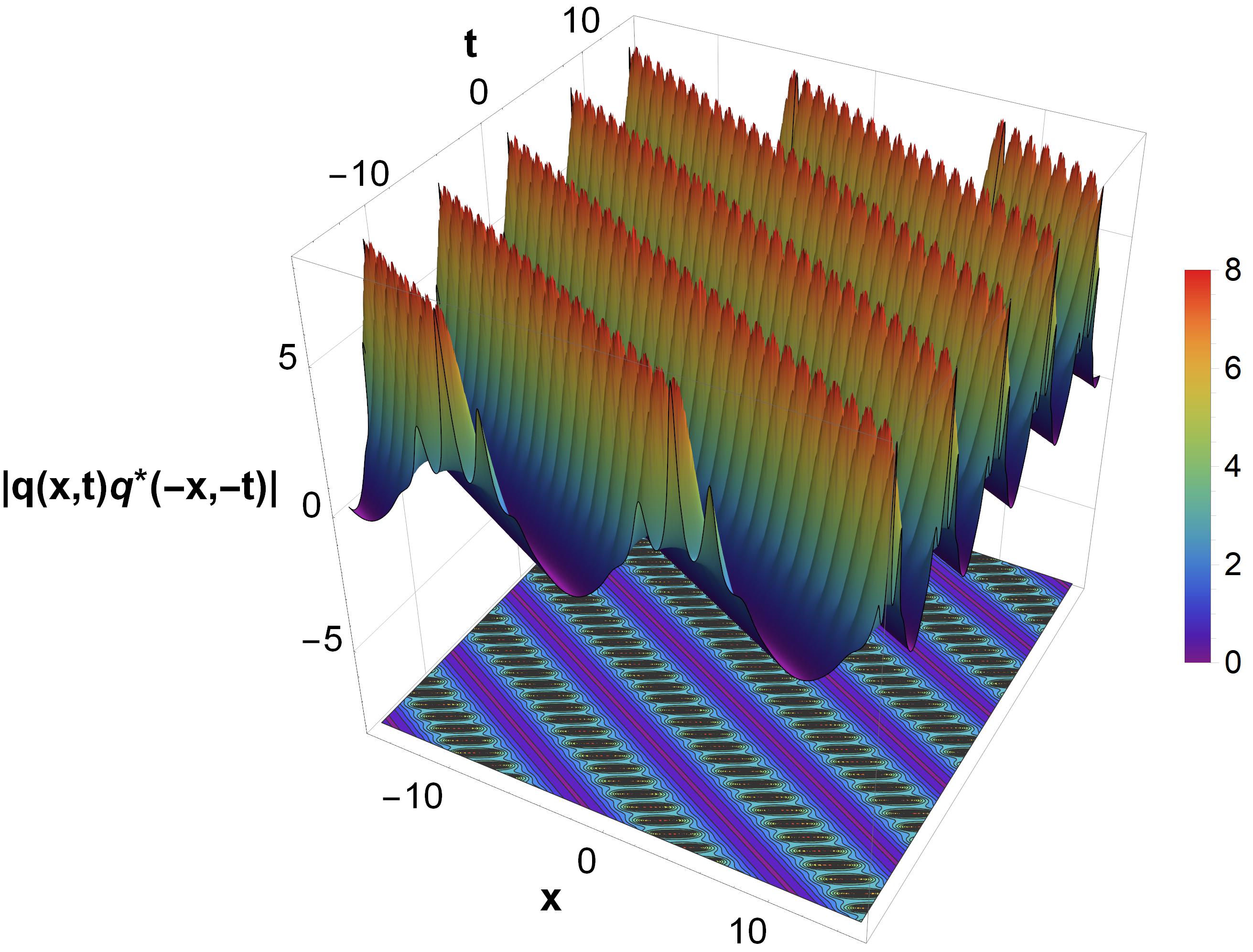}
		\caption{}
		\label{fig:2c}
	\end{subfigure}
	
	\vspace{0.3em}
	
	\begin{subfigure}{0.30\textwidth}
		\centering
		\includegraphics[width=\textwidth]{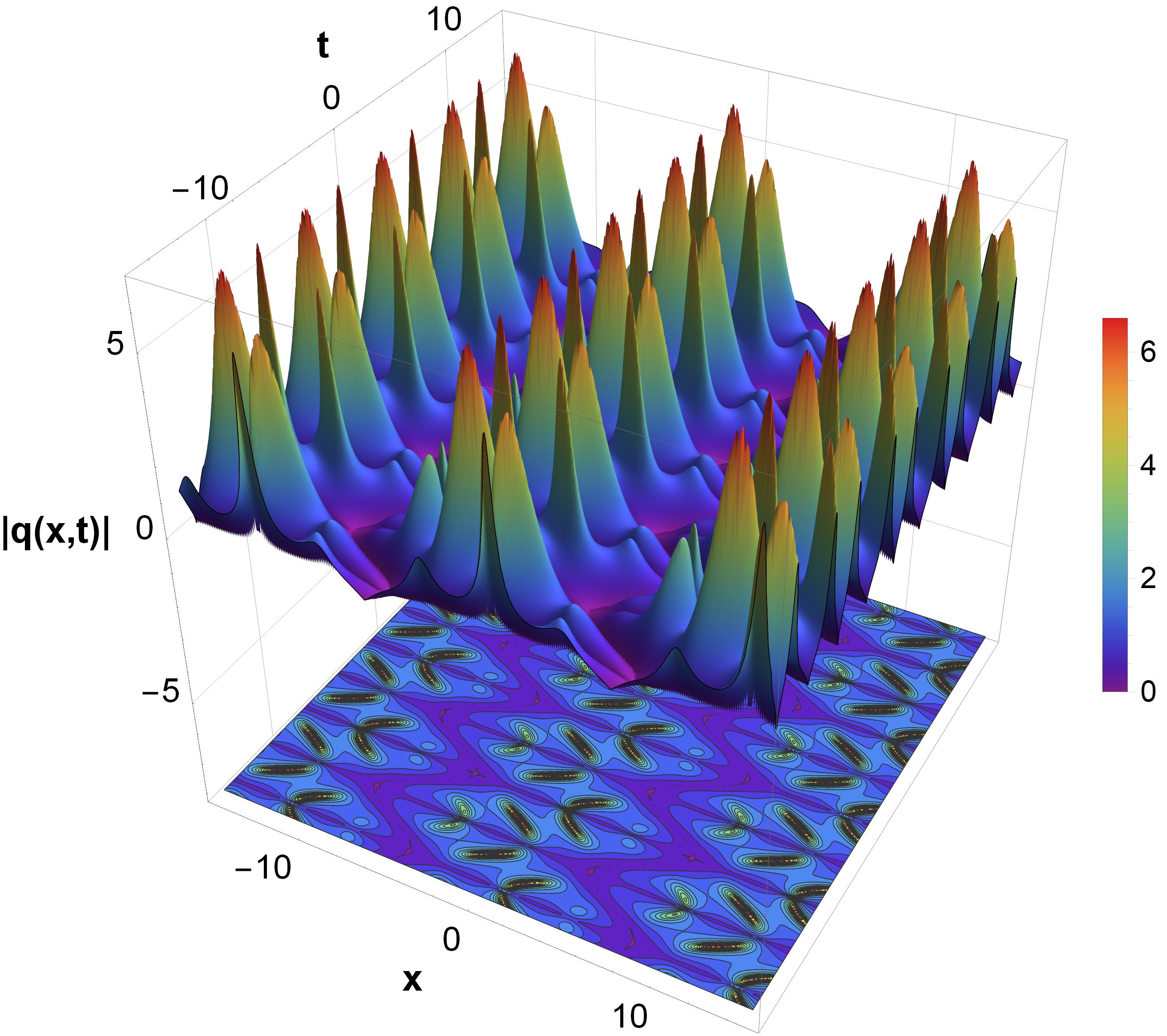}
		\caption{}
		\label{fig:2d}
	\end{subfigure}\hfill
	\begin{subfigure}{0.30\textwidth}
		\centering
		\includegraphics[width=\textwidth]{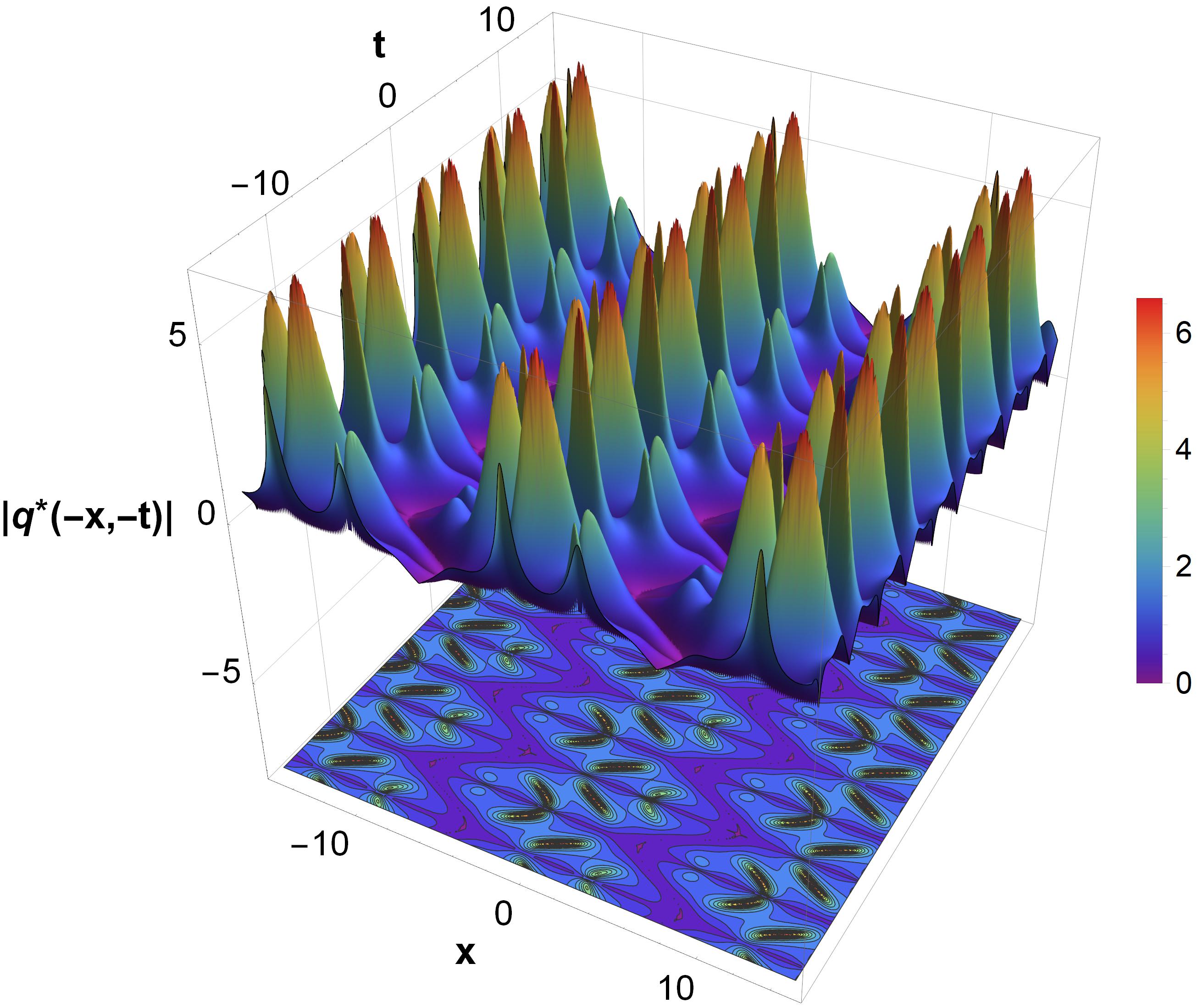}
		\caption{}
		\label{fig:2e}
	\end{subfigure}\hfill
	\begin{subfigure}{0.30\textwidth}
		\centering
		\includegraphics[width=\textwidth]{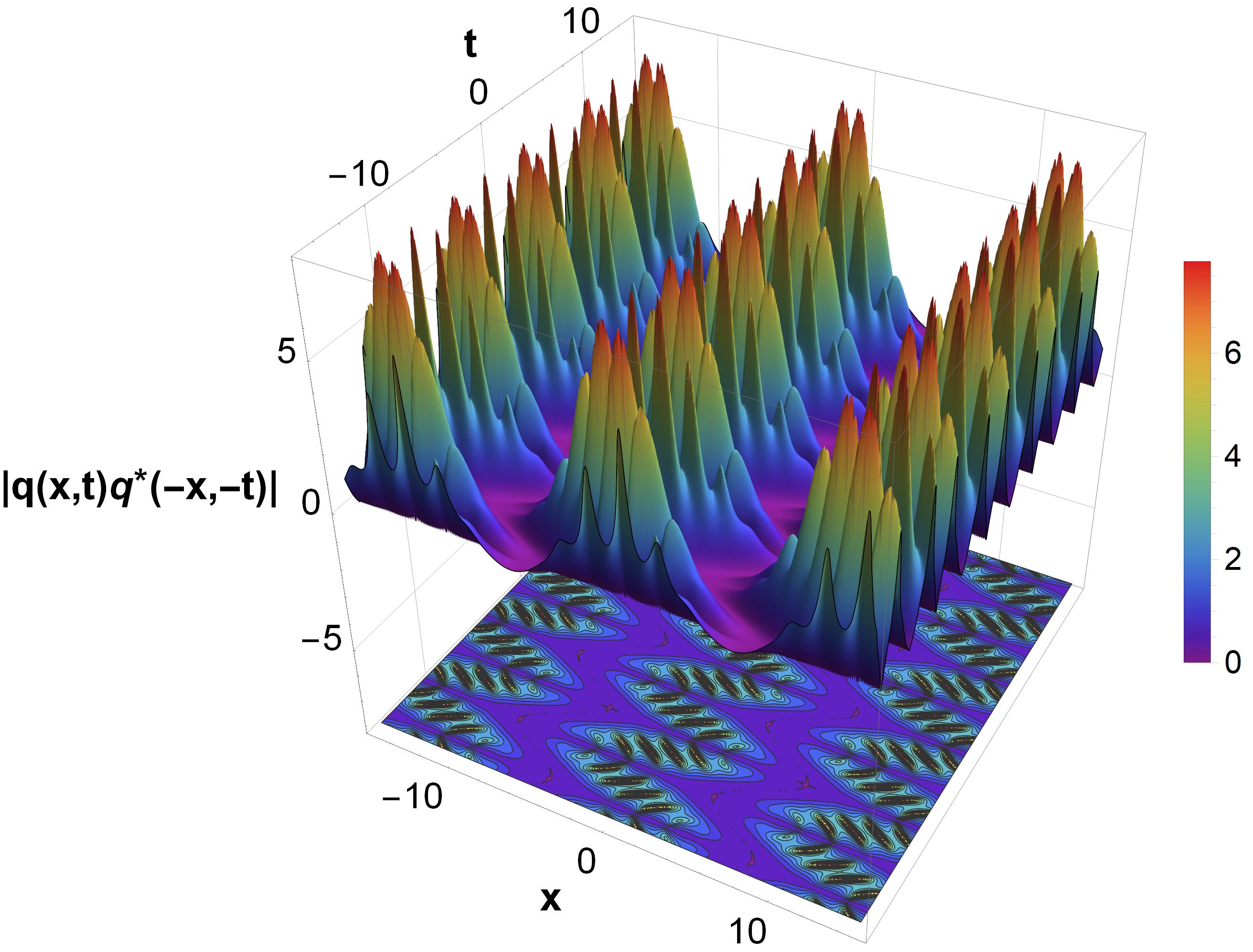}
		\caption{}
		\label{fig:2f}
	\end{subfigure}
	
	\vspace{0.3em}
	
	\begin{subfigure}{0.30\textwidth}
		\centering
		\includegraphics[width=\textwidth]{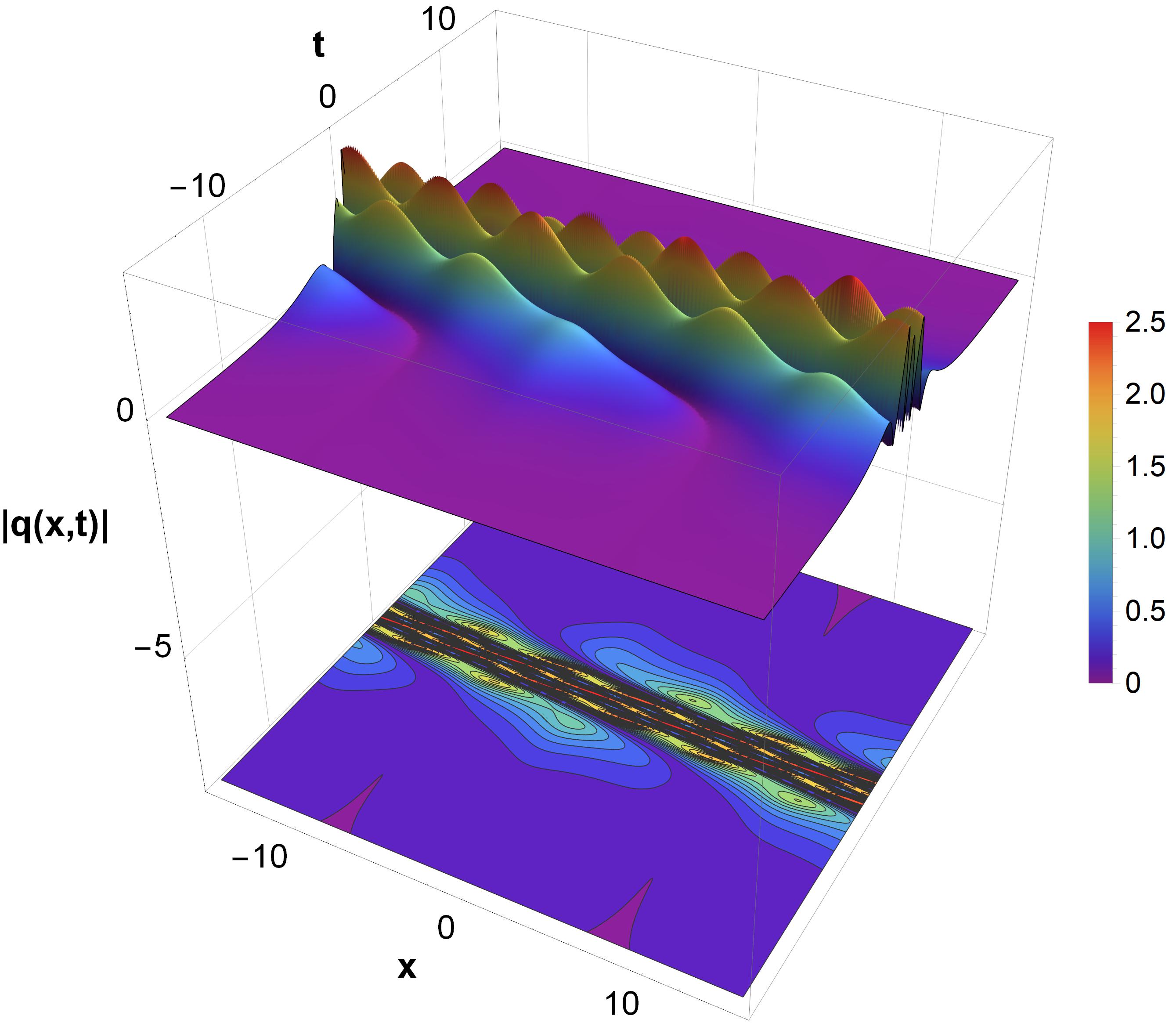}
		\caption{}
		\label{fig:2g}
	\end{subfigure}\hfill
	\begin{subfigure}{0.30\textwidth}
		\centering
		\includegraphics[width=\textwidth]{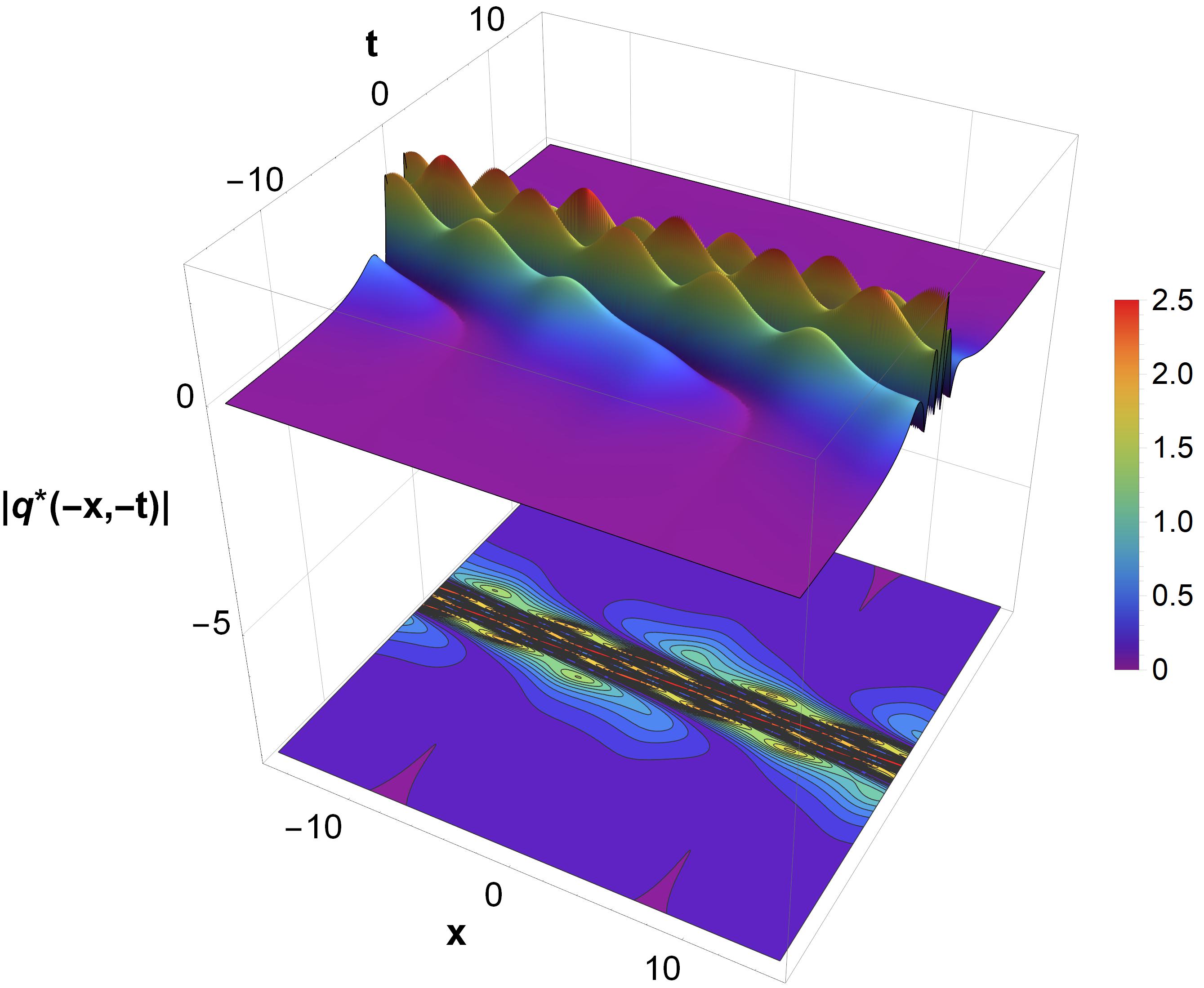}
		\caption{}
		\label{fig:2h}
	\end{subfigure}\hfill
	\begin{subfigure}{0.30\textwidth}
		\centering
		\includegraphics[width=\textwidth]{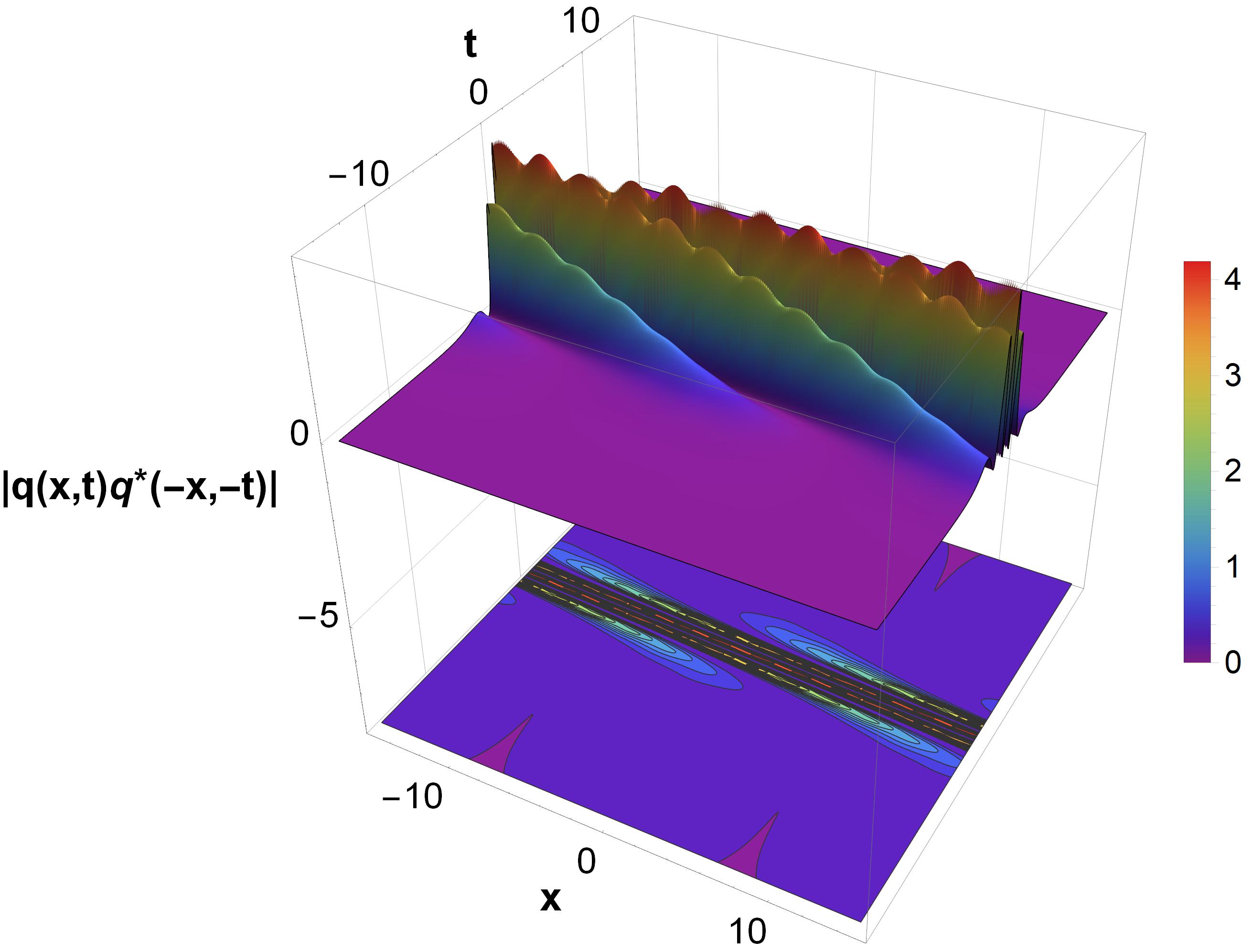}
		\caption{}
		\label{fig:2i}
	\end{subfigure}
	
	\vspace{0.3em}
	
	\caption{\scriptsize 3D plot of 2-SS given in Eqs. (\ref{sol3}) and (\ref{sol4}) for (a), (b), and (c) are combination of 
		periodic waves and breather like pattern obtained when $M(t)=1$, $N(t)=1$, 
		$\Gamma(t)=0$, $p_1=i$, $p_1^*=i$, $p_2=0.5i$, $p_2^*=0.5i$ and (d), (e), 
		and (f) are also periodic waves when $M(t)=cos t$, $N(t)=1$, 
		$\Gamma(t)=-\frac{1}{2}\tan t$, $p_1=i$, $p_1^*=i$, $p_2=0.5i$, 
		$p_2^*=0.5i$. In (g), (h), and (i) we show soliton like structure when 
		$M(t)=\text{sech}\, t$, $N(t)=1$, $\Gamma(t)=-\frac{1}{2}\tanh t$, 
		$p_1=0.1i$, $p_1^*=0.1i$, $p_2=0.5i$, $p_2^*=0.5i$. All the other 
		parameters are chosen as $\alpha_1=1$, $\alpha_1^*=1$, $\alpha_2=1$, 
		$\alpha_2^*=1$ and figures are obtained for $|q(x,t)|$, $|q^*(-x,-t)|$ 
		and $|q(x,t)q^*(-x,-t)|$, respectively.}
	\label{fig:two-solitons}
\end{figure}
	\begin{align*}
	\beta_1&=
	- \frac{p_1^2 \, p_1^* \, \alpha_1 \, \alpha_1^*}{(p_1 + p_1^*)^2 }, \quad
	\beta_1^*=
	\frac{{p_1^*}^2 \, p_1 \, \alpha_1 \, \alpha_1^*}{(p_1^* + p_1)^2 },\quad
	\beta_2=- \frac{p_1^2 \, p_2^* \, \alpha_1 \, \alpha_2^*}{(p_1 + p_2^*)^2 },\quad 
	\beta_2^*=\frac{(p_1^*)^2 \, p_2 \alpha_1^* \, \alpha_2}{(p_1^* + p_2)^2 },\\
	\beta_3&=- \frac{p_2^2 \, p_1^* \, \alpha_2 \, \alpha_1^*}{(p_2 + p_1^*)^2 }, \quad
	\beta_3^*=\frac{{p_2^*}^2 \, p_1 \, \alpha_2^* \, \alpha_1}{(p_2^* + p_1)^2 }, \quad
	\beta_4=- \frac{p_2^2 \, p_2^* \, \alpha_2 \, \alpha_2^*}{(p_2 + p_2^*)^2 },\quad 
	\beta_4^*=\frac{{p_2^*}^2 \, p_2 \, \alpha_2 \, \alpha_2^*}{(p_2^* + p_2)^2 },
\end{align*}	
\begin{align*}
	\beta_5&=\frac{p_2^2 \, p_2^* \, p_1^2 \, p_1^* \, \alpha_2 \, \alpha_2^* \, \alpha_1 \, \alpha_1^* \, (p_1 - p_2)^2 \, (p_1^* - p_2^*)^2}{(p_1 + p_1^*)^2 \, (p_1 + p_2^*)^2 \, (p_2 + p_1^*)^2 \, (p_2 + p_2^*)^2 },\\
	\beta_5^*&=	\frac{{p_2^*}^2 \, p_2 \, {p_1^*}^2 \, p_1 \, \alpha_2^* \, \alpha_2 \, \alpha_1^* \, \alpha_1 \, (p_1^* - p_2^*)^2 \, (p_1 - p_2)^2}{(p_1^* + p_1)^2 \, (p_1^* + p_2)^2 \, (p_2^* + p_1)^2 \, (p_2^* + p_2)^2 }
\end{align*}	
where $\alpha_1$, $\alpha_1^*$ $\alpha_2$, $\alpha_2^*$, $p_1$, $p_2$,  $p_1^*$, and $p_2^*$ are arbitrary distinct complex constants.\\
\indent The structures of (\ref{sol3}-\ref{sol4}) are shown in Figs. \ref{fig:two-solitons} . Under constant dispersion and nonlinearity, $M(t)=N(t)=1$, the profile combines periodic waves and breather-type patterns, as seen in Figs.\ref{fig:2a}-\ref{fig:2c} for $|q(x,t)|$, $|q^{*}(-x,-t)|$, and $|q(x,t)q^{*}(-x,-t)|$. Like the periodic waves in Figs \ref{fig:1a}-\ref{fig:1c}, these structures are also periodic. Following the explanation in the previous subsection, we first analyze with $M(t)=cos t$ where the dispersion varies with time. As shown in Figs. \ref{fig:2d}-\ref{fig:2f}, the solution again exhibits a periodic wave. However, the periodicity pattern differs from Figs. \ref{fig:1d}-\ref{fig:1f}. Next, with $M(t)=sech t$, the structure resembles a  {\color{blue}\textit{2}-SS} interaction, but a clear collision is not visible. The {\color{blue}\textit{1}-SS}-like profile obtained here resulted by modulating the periodic wave under constant dispersion via the $sech t$ dispersion function. Consequently, the shapes in Figs. \ref{fig:2a}-\ref{fig:2b} become significantly altered, resulting in a blurred or indistinct {\color{blue}\textit{2}-SS} pattern, as illustrated in Figs.{\ref{fig:2g}--\ref{fig:2i}.
	

	\subsection{N-bright soliton solution}
	Systematically, generalizing the above discussed method using the initial seed solution as $g_{1}=\sum_{n=1}^{N}\alpha_{n}e^{\theta_{n}}$, and $g^*_{1}=\sum_{n=1}^{N}\alpha^*_{n}e^{\theta^*_{n}}$, we provide the scheme to obtain the $N$- bright soliton solution of (\ref{cfle3})-(\ref{cfle4}) as,
	\begin{align}
		\label{Nsol}
		q(x,t)=\sqrt{\frac{M(t)}{N(t)}} \frac{ \sum_{n=1}^{N} \epsilon^{2n-1} g_{2n-1}}{1+\sum_{n=1}^{N} \epsilon^{2n} f_{2n}},\\ \label{Nsol2}
		q^*(-x,-t)=\sqrt{\frac{M(t)}{N(t)}} \frac{ \sum_{n=1}^{N} \epsilon^{2n-1} g^*_{2n-1}}{1+\sum_{n=1}^{N} \epsilon^{2n} f^*_{2n}}.	
	\end{align}
	Therefore, proceeding in a similar manner as in the cases of {\color{blue}\textit{1}-SS} ($N=1$) and {\color{blue}\textit{2}-SS} ($N=2$) discussed in the previous subsections, we can obtain the arbitrary $N$-soliton solution systematically.
	
	\section{Conclusion}
	In this letter, we have investigated a variable-coefficient FLE with gain and loss, reducing it to a nonlocal $\mathcal{PT}$-invariant vcFLE under certain symmetry conditions. As long as the coefficient functions $M(t)$ and $N(t)$ are even functions of time and $\Gamma(t)$ is an odd function, the system is $\mathcal{PT}$-invariant under certain conditions of dispersion, nonlinearity, and gain/loss coefficients. We construct the associated Lax pair and so proved that the system is integrable.
	Using a nonstandard Hirota bilinearization technique, we found explicit {\color{blue}\textit{1}-SS} and {\color{blue}\textit{2}-SS} as well as a systematic method for constructing $N$-soliton solutions. Our findings show that the nonlocal FLE has symmetry-preserving and symmetry-breaking solutions depending on the soliton parameters and coefficient functions. These findings reveal a richer dynamical landscape and demonstrate the nonlocal vcFLE's ability to support varied nonlinear structures. Future research may use these results to examine more generic nonlocal variable-coefficient integrable systems.

\backmatter

\bmhead{Acknowledgement}

S.T. wants to thank DST, Govt of India, for the INSPIRE Fellowship (Award No. DST/INSPIRE Fellowship/2020/IF200278).
R.R. would like to acknowledge the financial support in the form of the DST-ANRF National Postdoctoral Fellowship (File No. PDF/2023/001115).
M.L. acknowledges DST-ANRF, India, for the award of a DST-ANRF National Science Chair (NSC/2020/000029).

\section*{Statements \& Declarations}

\bmhead{Funding}
S.T. wants to thank DST, Govt of India, for the INSPIRE Fellowship (Award No. DST/INSPIRE Fellowship/2020/IF200278).
R.R. would like to acknowledge the financial support in the form of the DST-ANRF National Postdoctoral Fellowship (File No. PDF/2023/001115).
M.L. acknowledges DST-ANRF, India, for the award of a DST-ANRF National Science Chair (NSC/2020/000029).

\bmhead{Competing Interests}

The authors have no relevant financial or non-financial interests to disclose.

\bmhead{Authors Contribution}
Conceptualization: All the authors; Formal Analysis: Sagardeep Talukdar, R. Ramakrishnan; Supervision: Sudipta Nandy, M. Lakshmanan; Validation: Sudipta Nandy, M. Lakshmanan; Writing - original draft: Sagardeep Talukdar; Writing - review \& editing: All the authors.


\bibliography{references}

\end{document}